\documentclass[]{aa}  

\newcommand{\cd}{d$^{-1}$}
\newcommand{\kms}{km\,s$^{-1}$}

\newcommand{\vsini}{$v\sin{i}$}
\newcommand{\teff}{\ensuremath{T_{\rm{eff}}}}             
\newcommand{\logg}{\ensuremath{\log g}}                     

\usepackage{graphicx}
\usepackage{txfonts}
\usepackage{natbib}
\begin{document}
   \title{Spectroscopic monitoring of the Herbig Ae star HD\,104237  \thanks{Based on observations collected at the 1.9m SAAO Radcliffe telescope}}

   \subtitle{II.  Non-radial pulsations, mode analysis and fundamental stellar parameters}

 \author{%
 	 A. Fumel\inst{\ref{inst:irap1},\ref{inst:irap2}} \and
          T. B\"ohm\inst{\ref{inst:irap1},\ref{inst:irap2}} 
          }

   \institute{ 
              Universit\'e de Toulouse; UPS-OMP; IRAP; Toulouse, France\label{inst:irap1}\\
              \email{aurelie.fumel@ast.obs-mip.fr, boehm@obs-mip.fr}
       \and       
              CNRS; IRAP; 14, avenue Edouard Belin, F-31400 Toulouse, France\label{inst:irap2}
      }

   \date{Received August 3$^{rd}$, 2011; accepted XX, 2011}

  \abstract
   {Herbig Ae/Be stars are intermediate-mass pre-main sequence (PMS) stars showing signs of intense activity and strong stellar winds, whose origin is not yet understood in the frame of current  theoretical models of stellar evolution for young stars.  The evolutionary tracks of the earlier Herbig Ae stars cross for a significant fraction of their evolution towards the main sequence the theoretical PMS instability strip located roughly in the same area of the HR diagram as the $\delta$ Scuti variables. Many of these stars exhibit pulsations of $\delta$ Scuti type.}
   {We carry out  a thorough analysis of the line profile variations of the pulsating prototype Herbig Ae star HD\,104237 based on high-resolution spectroscopical time-series, identifiy the dominant non-radial pulsation mode, and perform the most accurate determination of its fundamental parameters in preparation of an asteroseismic modeling.}
   {HD\,104237 is a pulsating Herbig Ae star with eight detected frequencies based on the analysis of radial velocity variations. In this article, we reinvestigated an extensive high-resolution quasi-continuous spectroscopic data set in order to search for very faint indications of non-radial pulsations in the line profile. To do this,  we worked on dynamical spectra of equivalent photospheric (LSD) profiles of HD\,104237.  A 2D Fourier analysis (F2D) was performed of the entire profile and enabled the identification of the dominant non-radial pulsating mode. The temporal variation of the central depth of the line was studied with the time-series analysis tools \emph{Period04} and \emph{SigSpec}. The development of a highly accurate continuum normalization method enabled us to determine reliably the equivalent width of chosen absorption lines and to perform a new determination of the fundamental stellar parameters.}
   {Following the previous studies on this star, our analysis of the dynamical spectrum of recentered LSD profiles corresponding to the $22^{\rm nd}-25^{\rm th}$ of April 1999 nights spectra has confirmed the presence of multiple oscillation modes of low-degree $\ell$ in HD\,104237 and led to the first direct detection of a non-radial pulsation mode in this star: the dominant mode $F1$ was identified by the Fourier 2D method having a degree $\ell$ value comprised between 1 and 2, the symmetry of the pattern variation indicating an azimuthal order of $\pm$ 1. 
The detailed study of the fundamental stellar parameters has provided a $T_{\rm{eff}}$, log g and iron abundance of $8550\pm150$\,K, $3.9\pm0.3$ and $-4.38\pm0.19$\, (i.e. [Fe/H]$=+0.16\pm0.19$\,), respectively.}
   {}


   \maketitle


\section{Introduction}

Pre-main sequence (PMS) stars of intermediate mass (2 to 10 $\rm{M_{\odot}}$) are known as Herbig Ae/Be (hereafter HAeBe) stars, following a first classification by \cite{herbig1960}. As their name indicates, they are of spectral type A or B and luminosity classes III to V (for a review, see \citealt{waterswaelkens1998}). Their Spectral Energy Distribution (SED) is characterized by infrared (IR) excess and ultraviolet (UV) extinction mainly due to thermal re-emission from UV to IR of a circumstellar (CS) dust shell or disk (or both), making the photometric assessment of spectral types and stellar temperatures of these stars rather uncertain.

HAeBe stars show signs of intense stellar activity, variability and winds. Indeed, the frequently observed emission in  Mg II h and k lines,  Ca II IR triplet,  Ca II H and K line, He I 5876\,\AA, many Fe\,II lines, Na I D, N V, Si IV and C IV resonance lines, imply the high temperature of a chromosphere or corona to be formed. Short-term variability of many of these lines are observed and P Cygni profiles of H$\alpha$, H$\beta$, Mg II h and k lines present in the spectra of some of these stars indicate the presence of strong and structured stellar winds (see e.g. \citealt{praderie1982,praderie1986,catalatalavera1984,catala1986a,catala1986b,catalakunasz1987,catala1988,catala1993,bohmcatala1995,bohm1996}).

A wide-spread idea consists in invoking magnetism as being responsible for such active stellar phenomena, requiring either the presence of a primordial magnetic field or a dynamo mechanism whose basic ingredients are classically convection and rotation. 
On one hand, recent spectropolarimetric observations of Herbig stars indicate the presence of significant magnetic fields only in a small 10\% fraction of them \citep{wade2007}, these results being in agreement with the primordial fossil field hypothesis. On the other hand, considering a possible dynamo mechanism, the position of HAeBe stars in the HR diagram indicates that they are in the radiative phase of their quasi-static  contraction towards the main sequence \citep{iben1965,gilliland1986}. Therefore, in absence of subphotospheric convection, a classical solar-type magnetic dynamo mechanism can not be at work.

The HAeBe stars activity remains therefore quite paradoxical in the frame of current theoretical evolutionary models for PMS stars, although there exist some approaches to understand its origin. As possible explanations, \cite{pallastahler1990} suggested that outer an convection zone could be induced by deuterium burning. \cite{vigneron1990}  proposed the internal rotation as energy source, the friction exerted at the stellar surface by the angular momentum losses creating a non-convective turbulent layer, responsible for a dynamo effect; this model was later refined by \cite{lignieres1996}. 

An external origin, involving an accretion disk dissipating gravitational energy into a boundary layer at the stellar surface,  was invoked by \cite{bertout1988} to explain activity in classical TTauri stars, TTauri stars being the lower mass PMS counterparts to HAeBes.  However, \cite{bohmcatala1995} showed that the energy flux emitted in winds and some activity tracers seems to increase with effective temperature, which is in favor of an internal activity source rather than a CS origin.
Thus, as of today, growing evidences tend to indicate that the energy needed to produce this activity might be of internal stellar origin, but no definite answer has been provided. It is a major concern for testing young stellar evolutionary theory to solve this still open question about HAeBe stellar activity, by constraining the internal structure of these objects using asteroseismic techniques, i.e. the analysis and modeling of stellar pulsations, if observed.

Photometric and spectroscopic variability have indeed been observed in HAeBe stars, on very different time scales (from minutes to years).
Large and irregular photometric variations ($\geq 0^{\rm{m}}.5$) with time scales from weeks to years are thought to be due to variable obscuration by CS patchy dust clouds, particularly strongly in the younger objects \citep{vdancker1998}. Variations of the order of a tenth of magnitude (mag) in amplitude, with time scales of hours to days, can be explained as due to clumpy accretion (\citealt{vdancker1998} and references therein) or due to chromospheric activity, including winds, modulated by the star's rotation \citep{catala1999}. Stellar pulsations, which are here the subject of our interest, induce typically a millimagnitude (mmag) photometrical variability with time scales of minutes to hours.

The existence of such stellar pulsations in HAeBe stars are known since \cite{breger1972} discovered two pulsating candidates, V588~Mon and V589~Mon, in the young open cluster NGC~2264. Subsequently, two additional pulsating Herbig stars were detected: HR~5999 \citep{kurtzmarang1995} and HD\,104237 \citep{donati1997}.
Motivated by these detections, \cite{marconipalla1998} investigated theoretically a PMS instability strip for the first three radial modes of pulsations and concluded on a topology of this instability strip, corresponding roughly to the same area in the HR diagram as the $\delta$ Scuti variables. The study of \cite{marconipalla1998} showed that most of the Herbig Ae stars are expected to cross it for a significant fraction of their evolution to the main sequence (5 to 10\% of their PMS phase). Since the precise location and boundaries of the PSM instability strip has not yet been observationnally constrained, it is necessary to identify and study the largest number of PMS pulsating objects, giving priority to multiperidodic pulsators which are suitable and promising candidates for future asteroseismic modeling, and, by this mean, supplying constraints on theoretical evolutionary pre-main sequence models.

As of today,  a significant number of Herbig Ae stars, either field stars or members of young open clusters, have indeed revealed to be pulsating at timescale typical of $\delta$ Scuti stars, i.e. with short periods (from $\approx$ 20 min to several hours) and small amplitudes (from mmag to few hundreths of mag in case of photometry, and less than typically 1 - 2 \kms for radial velocity studies) (for reviews see e.g. \citealt{catala2003,marconipalla2004,zwintz2004,zwintz2008a}). Most of them have been studied in photometry, but only very few in spectroscopy (e.g. \citealt{bohm2004,bohm2009}). However, due to cancellation effects, only modes of low degree are detected in photometry or from radial velocity curves.  Therefore, high-resolution spectroscopy is necessary to study the weak variations induced by non-radial pulsations in the rotationally-broadened photospheric line profiles, and thus to carry out a comprehensive pulsational analysis in preparation for seismic modeling. 
It should be noted that  binarity is common amongst HAeBe stars with approximately 60$\%$ \citep{baines2006}. 
As we will see in Sect. \ref{previous}, HD\,104237 is also a spectroscopic binary, which complicates tremendously the analysis of its spectrum.

This article is the second of a series on the prototype Herbig Ae star HD\,104237, starting with the article by \cite{bohm2004}.
The first article presented the detection of multiperiodic oscillations in the radial velocity curves of the star and the analysis of its binary orbit, amongst others. In this present article, we describe the direct search and detection of oscillations in the line profiles, the mode identification of the dominant non-radial pulsation mode, as well as a fundamental stellar parameter determination by spectroscopic means. This work represents therefore the basis for the asteroseismic modeling we will present in a forthcoming article \citep{fumel2011}.

The article is structured as follows: in Sect. \ref{previous}, we review in detail previous works on HD\,104237, Sect. \ref{obs} sums up the spectroscopic observations we worked with and data reduction, Sect. \ref{nonrad} presents our detection and identification of non-radial pulsations in HD\,104237, Sect. \ref{norm} describes the spectral continuum determination tool we have developed to obtain optimized high-quality spectra, in Sect. \ref{funda} we detail the fundamental parameter determination of HD\,104237 we carried out and finally we conclude and discuss our results in Sect. \ref{conc}.

\section{Previous works on HD\,104237}
\label{previous}

\cite{hu1989,hu1991} derived from photometric and spectroscopic studies that HD\,104237, or DX Cha, is a member of the Herbig Ae/Be group. Moreover, it is a prototype of this group of stars and a particularly suitable target since it is very bright ($m_{\rm{v}} = 6.6$), which enables to observe it in high resolution spectroscopic mode. It is located in the $\epsilon$ Chameleontis young stellar group at a \emph{Hipparcos\/} distance of $116 \pm 8$ pc \citep{vdancker1997}.
HD\,104237 is actually a multiple system whose primary component (hereafter simply called HD\,104237 or the "primary") is a Herbig Ae star with several low-mass companions at separations between 1'' and 15'' \citep{feigelson2003}, including a close K3 companion in an eccentric 19.9-day orbit (hereafter HD\,104237b or the "secondary"). This very close component of approximate spectral class K3 forms with the primary a spectroscopic binary \citep{bohm2004}. Indeed, K3-type spectral features (including Li I 6707\,\AA~ and Ca I 6718\,\AA\ lines) are observed in the primary spectra \citep{feigelson2003}, that makes its spectroscopic analysis trickier since one has to take into account the pollution generated by the faint secondary component in the dominant primary spectrum.

\subsection{Previous fundamental parameter determination}
The strong IR excesses and the UV-extinction due to the presence of CS dust , the emission components in many absorption lines, the P Cygni profiles observed notably in the Balmer lines and the spectroscopic binary pollution make the determination of most fundamental stellar parameters of HD\,104237 difficult. Thus, the spectral types avalaible in the literature for HD\,104237 range from A0 to A8 (implying a large range in possible $T_{\rm{eff}}$ values) and its bolometric luminosity from $\approx20\ \rm{to}\ 60\ L_{\odot}$. The associated stellar parameters such as the mass and the age of the star can be estimated from temperature and luminosity with stellar evolutionary models. They rely therefore strongly both on the accuracy of the location of the star in the HR diagram and on the adopted evolutionary models and are also not precisely determined, as well as all the parameters that result from them (e.g. log $g$). The most important parameters of HD\,104237 (primary and secondary) found in the literature are chronologically summarized in Table~\ref{table1} and are detailed below.

\cite{hu1989} based their first determination of a A0Vpe spectral type on the absorption profiles of the Balmer lines and the strong Ca II K line (3933.7\,\AA). Subsequently, the presence of emission components in the H${\alpha}$ and H${\beta}$ was understood to be misleading the determination of the spectral type. In a second article \cite{hu1991} therefore derived an A4Ve spectral type from UV absorption lines. From an integrated radiant flux (0.1 to 10 $\mu$m) of $8.62 \times 10^{-8}$ erg cm$^{-2}$ s$^{-1}$ and a distance of 88 pc they inferred a ${\rm{log}}\ (L_{\star}/L_{\odot})$ of 1.34. Using the dwarf temperature scale of \cite{schmidtkaler1982}, they converted this spectral type to a $T_{\rm{eff}}$ of 8450\,K from which they obtained a stellar radius of 2.2 R$_{\odot}$. \cite{brown1997} carried out a study of the hot disk wind of HD\,104237 by means of UV spectra and assumed it to be of spectral type A7IVe. Using the accurate absolute astrometric and photometric data yielded by the Hipparcos satellite, \cite{vdancker1997} computed new $T_{\rm{eff}}$, ${\rm{log}}\ (L_{\star}/L_{\odot})$, mass and age values for HD\,104237 (cf. Table~\ref{table1}). 
Although their method included corrections taking into account the fact that HAe/Be stars often exhibit anomalous extinction law for their circumstellar material and UV spectra often lead to overestimation of temperatures due to the presence of heated layers in the immediate surrounding of the star's p hotosphere, they used as a spectral type the non-corrected result of \cite{hu1989} based on optical spectra, whose many lines are polluted by emission components. In a next article \cite{vdancker1998} proposed a cooler spectral type (A4IVe), in agreement with \cite{hu1991}, and derived corresponding new parameters. The uncertainties on the effective temperature were estimated values and not determined in a statistical procedure.

In their investigation of the environment of HD\,104237, \cite{grady2004} derived a new spectral type of A7.5-A8Ve based on the comparison between FUV and UV spectra of HD\,104237 and others stars with well-known spectral types. Again, the error bars on the spectral type were empirically determined. This cooler spectral type, close to the determination of \cite{brown1997}, and the ${\rm{log}}\ (L_{\star}/L_{\odot})$ derived by the authors led to an age of 5 Myr, which is noticeably greater than the previous age estimations \citep{vdancker1997,vdancker1998,feigelson2003} but closer to the ages of the HD\,104237 quintet low-mass members \citep{feigelson2003}. Using evolutionary tracks and isochrones from \cite{pallastahler2001} and both the stellar parameters from \cite{vdancker1998} and \cite{grady2004}, \cite{bohm2004} concluded on an age of 2 Myr for HD\,104237, in agreement with \cite{vdancker1997,vdancker1998} and \cite{feigelson2003}, and independently concluded on a similar value of the stellar mass  $M_{\star}=2.2\pm0.1M_{\odot}$ as previous studies. Moreover they calculated a mass ratio $M_{P}/M_{S}$ of $1.29\pm0.02$, based on measures of the primary and secondary radial velocities with respect to the systemic velocity of the binary system, that enabled them to estimate the mass of the secondary to be of $1.7\pm0.1M_{\odot}$. Assuming the same age for both components, they confirmed a K3 spectral type for the secondary and assessed a temperature of about 4750\,K and a luminosity of one tenth of $L_{\rm{P}}$. Starting from the spectral type proposed by \cite{grady2004}, \cite{luhman2004} computed a new $T_{\rm{eff}}$ (with \citealt{schmidtkaler1982} temperature scale) and ${\rm{log}}\ (L_{\star}/L_{\odot})$ (from $I$-band magnitudes). Placing these new parameters in a HR diagram with evolutionary tracks from \cite{pallastahler1999} enabled \cite{luhman2004} to confirm an age of about 2 Myr. Finally, based on low-resolution spectra and going by the fact that the strengths of Ca II and H$\delta$ lines undergo a reverse relationship with temperature, \cite{lyo2008} managed to constrain the spectral type of HD\,104237, even in case of contamination by emission of these lines. Since they obtain a spectral type of A4 from the Ca II line and A5 from the H$\delta$ line, they adopted a spectral type of A4/A5 for this star. 

\cite{ackewaelkens2004} used the $T_{\rm{eff}}$ and log $g$ presented in \cite{meeus2001} (and references therein) to carry out a chemical analysis of several PMS stars including HD\,104237. They measured the equivalent widths of many chemical elements and converted them to abundances using the program MOOG \citep{sneden1973}, but taking into account neither the binarity of this star (which adds a continuum veiling along with additional lines from the secondary) nor the asymmetries or the emission components in the line profiles. Using the solar abundances by \cite{andersgrevesse1989} they found a solar-like global metallicity ([M/H]$=+0.06\pm0.05$) and a solar abundances for several chemical elements including Fe ([Fe/H]=$+0.09\pm0.19$ where [Fe/H]=$\log{\left(\frac{N_{{\rm{Fe}}}}{N_{{\rm{H}}}}\right)_{{\rm{HD}}\,104237}}-\log{\left(\frac{N_{{\rm{Fe}}}}{N_{{\rm{H}}}}\right)_{\odot}}$).

\cite{donati1997} measured a $v\rm{sin}i$ of $12\pm2$ km s$^{-1}$, which suggests either that HD\,104237 is viewed close to pole-on or that  it is a moderate rotator. \cite{grady2004} found indeed an inclination of the stellar rotation axis of $18^{\circ}$$^{+14}_{-11}$, which implies an equatorial velocity of 38 \kms. \cite{bohm2006} detected a modulation of the H$\alpha$ line profile with a period of $100\pm5$ hrs and suggested a rotational origin,  yielding a similar inclination of $23^{\circ }$$^{+9}_{-8}$. This value of $P_{\rm{rot}}$ was confirmed by \cite{testa2008} in a X-ray study of the HD\,104237 system.

Concerning magnetism, \cite{donati1997} detected a weak field of about 50\,G in HD\,104237, but a later survey by \cite{wade2007} could not confirm it at this stage.

As can be seen from all the different contributions cited here-above, no in-depth spectral type determination has been done as of today: indeed, all determinations were based either on photometry or on low-resolution spectroscopy (except the rough estimate by  \citealt{donati1997}), and a thorough redetermination of fundamental parameters based on high-resolution spectroscopy is required for an independent conclusion. This is one of the major goals of the work presented in this article in Sect.  \ref{funda} and a basic requirement for a subsequent asteroseismic modeling  \citep{fumel2011}, but also in order to precise the PMS instability strip features and boundaries. Knowing these parameters will eventually allow us to better understand the origin of the activity observed in the Herbig Ae/Be stars and test the current theoretical evolutionnary models for PMS stars.

\begin{table*}[!ht]
\centering
\caption{Stellar parameters of HD\,104237. $2^{nd}$ column: P corresponds to the stellar parameters of  the primary and S to those of the secondary. Values in italics (log $g$ and $R/R_{\odot}$): computed from universal gravitation and Stefan-Boltzmann laws. References: [1] \cite{hu1989}, [2] \cite{hu1991}, [3] \cite{brown1997}, [4] \cite{vdancker1997}, [5] \cite{donati1997}, [6] \cite{vdancker1998}, [7] \cite{grady2004}, [8] \cite{bohm2004}, [9] \cite{luhman2004}, [10] \cite{ackewaelkens2004}, [11] \cite{lyo2008}, [12] \cite{bohm2006}. Mass and ages are determined using the model isochrones and isomass of: $^{\rm{(a)}}$ \cite{iben1984}, $^{\rm{(b)}}$ \cite{pallastahler1993}, $^{\rm{(c)}}$ \cite{siess2000}, $^{\rm{(d)}}$ \cite{pallastahler1999}, $^{\rm{(e)}}$ \cite{pallastahler2001}.} \label{table1}
\begin{tabular}{c|c|c|cc|c|c|c|c|c|c|c}
\hline
 Ref. &  & Sp. Type & ${\rm{log}}\ T_{\rm{eff}}$ & $T_{\rm{eff}}$ \tiny{(K)} & ${\rm{log}}\ (L_{\star}/L_{\odot})$ & log $g$ & $M_{\star}/M_{\odot}$ & $R_{\star}/R_{\odot}$ & Age \tiny{(Myr)} & $v{\rm{sin}}i\ (\rm{km.s^{-1})}$ & $i\ (^{\circ})$ \\
 \hline \hline
 [1] & P & A0Vpe & & & & & & & & & \\  \hline
 [2] & P & A4Ve & 3.93 & 8450 & 1.34 & \emph{4.1} & 2.1$^{\rm{(a)}}$ & 2.2 & & & \\  \hline
 [3] & P & A7IVe& & & & & & & & & \\  \hline
 [4] & P & A0Vpe [1] & $3.98\pm0.05$ & $9550^{\pm 550}$ & $1.77\pm0.06$ & \emph{3.9} & $2.5\pm0.1$$^{\rm{(b)}}$ & \emph{2.9} & $2.0\pm0.5$$^{\rm{(b)}}$ & & \\  \hline
 [5] & P & A4V & & & & & & & & $12\pm2$ & \\ \hline
 [6] & P & A4IVe$+$sh & $3.93\pm0.05$ & $8500^{\pm 500}$ & $1.55^{+0.06}_{-0.05}$ & \emph{3.9}  & 2.3$^{\rm{(b)}}$ & \emph{2.8} & 2.0$^{\rm{(b)}}$ & & \\ \hline
 [7] & P & A7.5-8Ve & 3.86 & $\approx7300$ & $1.42^{+0.04}_{-0.07}$ & \emph{3.7} & 2.1$^{\rm{(c)}}$ & \emph{3.3} & 5$^{\rm{(c)}}$ & & $18¡^{+14}_{-11}$ \\ \hline
 [8] & P & A7.5-8Ve [7] & & $\approx7300$ [7] & 1.42 [7] & \emph{3.7} & $2.2\pm0.1$$^{\rm{(e)}}$ & \emph{3.3} & 2$^{\rm{(e)}}$ & & \\
         & S & K3 & 3.675 & 4730 & 0.42 & \emph{3.9} & $1.7\pm0.1$$^{\rm{(e)}}$ & \emph{2.5} & 2$^{\rm{(e)}}$ & & \\ \hline
 [9] & P & A7.75 [7] & 3.88 & 7648 & 1.46 & \emph{3.8} & 2.2$^{\rm{(d)}}$ & \emph{3.1} & $\approx2$$^{\rm{(d)}}$ & & \\ \hline
 [10] & P & & 3.90 & 8000 & & 4.5 & & & & $10\pm1$ & \\ \hline
 [11] & P & A4-5Ve & & & & & & & & & \\ \hline
 [12] & P &  & & & & & & & & & $23^{+9}_{-8}$ \\
\hline
\end{tabular}
\end{table*}

\subsection{Previous pulsational analysis}

The primary component of the HD\,104237 system has revealed being a pulsating Herbig Ae star. From spectropolarimetric observations of active stars including HD\,104237, \cite{donati1997} detected for the first time radial velocity variations in equivalent photospheric LSD (cf. Sect. \ref{obs}) Stokes I spectra of this star and determined a period of 37.3\,min. This short period was incompatible with a rotational modulation or a secondary companion and was attributed to stellar pulsations. This pulsational variability, typical of $\delta$ Scuti-type variables, was confirmed with photometric measurements by \cite{kurtzmuller1999} who detected two frequencies. They computed the ratio of radial velocity to $V$ light amplitude and found values typical for a $\delta$ Scuti star. Moreover, they calculated a value for the pulsation constant $Q$ lower than the typical values given by the fundamental, the first and the second overtone in $\delta$ Scuti stars, indicating pulsations of higher overtone.

\cite{bohm2004} obtained in 1999 and 2000 an extensive high-resolution (R$\approx$35000) spectroscopic data set of HD\,104237 at the 1.9m Radcliffe telescope of the SAAO (see Sect. \ref{obs}). The resulting high quality radial velocity curve allowed them to detect for the first time by spectroscopic means multi-periodic oscillations in a PMS star (8 frequencies in the 1999 data set and 5 in the 2000 data set). Five frenquencies of 1999 were identified with frequencies of 2000, between 28.50 and 35.60 d$^{-1}$. No pulsation was detected in HD\,104237b in the night around the periastron, when both components are well separated in velocity, a result which is not astonishing given the position of the star in the HR diagram. The frequency differences of the 8, respectively 5 frequencies detected by \cite{bohm2004} were smaller thant the estimated large separation ($\Delta \nu_{0}\approx40-45\ \mu$Hz), leading the authors to conclude on the nonradial nature  of at least some of the pulsation modes detected in their work. More generally, no regular frequency pattern has been clearly identified in the pulsational behaviour of HD\,104237. \cite{bohm2004} pointed out (based on the frequency separation) the potential existence of 1 to 3 radial pulsation modes among the 5 frequencies observed both in 1999 and 2000, but at this time no clear mode identification in HD\,104237 could have been performed for none of the modes.
Since this first article was restricted to the analysis of radial velocity variations, we wanted to go a step further, and reanalysed the spectral line profiles themselves in the same data set; the result of this study is presented thereafter.
 
The previously determined pulsation characteristics are summarized in Table~\ref{table2}. The three last frequencies of 1999 from \cite{bohm2004} ($f_{6}$ to $f_{8}$) are included, although these authors considered them as less certain.

Using the results of \cite{bohm2004}, \cite{dupret2006,dupret2007} found in a preliminary seismic modeling of HD\,104237 that it is difficult to properly fit the observed frequencies with theoretical frequencies computed from models with $T_{\rm{eff}}$ and log($L/L_{\odot}$) presented in  \cite{vdancker1998} and \cite{grady2004}. An important result was their conclusion that HD\,1043237 is in fact not a typical $\delta$ Scuti type pulsator. Indeed, the p modes order corresponding to the observed frequencies are too high to be excited by the standard 
excitation mechanism of $\delta$ Scuti type variables, namely a $\kappa$-driving in the He II partial ionization zone. Such a modeling requires a precise knowledge of the stellar fundamental parameters which enter the evolution and oscillation codes and a good idea of $\ell$ and $m$ for each observed mode. However, both are poorly known or not known at all in the case of HD\,104237. In order to go further in this seismic study and to better constrain the internal structure of HD\,104237, we will propose a more thorough and complete modeling of HD\,104237 in a forthcoming article \citep{fumel2011}, using the accurate stellar parameters determination as well as the pulsational information described in the present article.

\begin{table}[h]
\caption{Previous pulsational analysis results for HD\,104237. Col. 5: references: (1) \cite{donati1997}, (2) \cite{kurtzmuller1999}, (3) \cite{bohm2004}.} \label{table2}
\begin{tabular}{c|l|l|l|c}
\hline
 Year of & Frequency & Period & Amplitude & Ref. \\
 observation & (d$^{-1}$) & (min) & & \\
 \hline \hline
 1993 & $f_{1}=37.3$   & $38.6\pm1$ & 0.65 \kms & (1) \\ \hline
 1995 & $f_{1}=39.6$   & $36.4\pm1$ & 0.65 \kms & (1) \\ \hline
 1998 & $f_{1}=33.29$ & 43.3 & $11.2\pm0.5$ mmag & (2) \\
           & $f_{2}=36.61$ & 39.3 & $3.4\pm0.5$ mmag &  \\ \hline
 1999 & $f_{1}=33.289$ & 43.257 & 1.320 \kms & (3) \\
           & $f_{2}=35.606$ & 40.443 & 0.474 \kms & \\
           & $f_{3}=28.503$ & 50.521 & 0.195 \kms & \\
           & $f_{4}=30.954$ & 46.521& 0.139 \kms & \\
           & $f_{5}=33.862$ & 42.525 & 0.099 \kms & \\
           & $f_{6}=32.616$ & 44.150 & 0.105 \kms & \\
           & $f_{7}=34.88$ & 41.28 & 0.1 \kms & \\
           & $f_{8}=35.28$ & 40.82 & 0.05 \kms & \\
            \hline
 2000 & $f_{1}=35.609$ & 40.439 & 0.328 \kms & (3) \\
           & $f_{2}=33.283$ & 43.265 & 0.258 \kms & \\
           & $f_{3}=31.012$ & 46.434 & 0.177 \kms & \\
           & $f_{4}=28.521$ & 50.489 & 0.165 \kms & \\
           & $f_{5}=32.375$ & 44.479 & 0.113 \kms & \\
\hline
\end{tabular}
\end{table}

\section{SAAO spectroscopic observations and data reduction}
\label{obs}

The analysis presented in this article is based on parts of the data set described in \citet{bohm2004}, namely the high-resolution spectroscopic observations obtained in April 1999 and April 2000 at the 1.9 m SAAO (South African Astronomical Observatory) Radcliff telescope; all data were obtained with the GIRAFFE fiber-fed echelle spectrograph with a resolving power of about $R \approx 35000$.

HD\,104237 was monitored in quasi-continuous mode throughout all the nights of the 1999 and 2000 runs (resp. 7 and 14 nights) in order to obtain high-resolution spectroscopic time-series necessary for an asteroseismic analysis.
The covered wavelength domain usually was 426-688 nm (spread over 50 orders) in 1999 and 436-688 nm (spread over 47 orders) in 2000. In order to sample properly the previously known main pulsation period of about 40 minutes, subexposure times of 5 minutes were needed, yielding about 8 spectra per pulsation period,  the brightness of HD\,104237 enabling such short exposure times. A standard calibration strategy was applied. Values of Signal to Noise Ratios (SNR) per pixel at 550 nm ranged from 30 to 120 in 1999 and from 30 to 110 in 2000, with typical values of 50-70. Most of the data reduction was carried out following standard reduction procedures using the "ESPRIT" spectroscopic reduction package \citep{donati1997}. Moreover, as the analysis of pulsations by high-resolution spectroscopy requires significant SNR, the least-square deconvolution (LSD) method as described in \citet{donati1997} was applied. This method makes use of the multiplex gain of more than 500 lines present in the spectrum of this star, providing a high SNR equivalent photospheric profile. More precisely, it assumes that the local profile of all selected spectral lines is similar in shape (with different individual central depths), the line parameters being extracted from an appropriate catalogue, corresponding as close as possible to the fundamental parameters of the star of interest known at this stage of data reduction as A4-type (based on the results announced in  \citealt{donati1997}). In a subsequent step, the more than 100 narrow telluric vapor lines present in the spectra yielded the necessary multiplex information for a high precision radial velocity correction. All spectra were shifted to the heliocentric rest wavelength, and calibrated to a final intrinsic precision estimated to be around $100\ \rm{ms^{-1}}$ for this data set. In addition, HJD (Heliocentric Julian Date) have been calculated for each LSD profile. Details of observations and data reduction are thoroughly described in \citet{bohm2004}.

The detection and identification of non-radial pulsations described in Sect. \ref{nonrad} has been carried out on the data collected during the SAAO observing run in 1999, where we especially concentrated our study on the data of the nights from $22^{\rm nd}$ to $26^{\rm{th}}$ of April,
these nights having yielded the longest individual time coverages. During the night of April 25$^{\rm th}$ 156 high-resolution spectra were obtained during 10.9 hrs.

The fundamental parameter determination described in Sect. \ref{funda} has been performed on data collected during the night of $12^{\rm{th}}$ of April 2000 at SAAO; on this date, the eccentric binary orbit with a period of 19.859 days (see  \citealt{bohm2004}) separates during periastron both spectra with a relative shift of 60 \kms, thus avoiding at maximum pollution of the primary spectrum by the much fainter secondary companion HD\,104237b.

The log of the observations we worked with are summarized in Table~\ref{tablog}.

\begin{table}[!h]
\centering
\caption{Log of the observations we worked with.} \label{tablog}
\begin{tabular}{c|c|c|c}
\hline
Date & Time series & Number of & SNR (pixel$^{-1}$) \\
         & duration (hrs) & spectra    & at 550 nm \\
\hline \hline
22 Apr. 1999 & 8.2 & 74 & 55-120 \\
23 Apr. 1999 & 10.2 & 91 & 40-90 \\
24 Apr. 1999 & 10.8 & 136 & 30-100 \\
25 Apr. 1999 & 10.9 & 156 & 30-60 \\
26 Apr. 1999 & 7.7 & 58 & 30-80 \\ \hline
12 Apr. 2000 & 3.6 & 35 & 40-60 \\
\hline
\end{tabular}
\end{table}

\section{Detection and identification of non-radial pulsations}
\label{nonrad}

Stellar pulsations are characterized by their frequencies and corresponding modes, specified by three quantum numbers: the order $n$, related to the number of radial nodes, the degree of the mode $\ell$, indicating the total number of surface nodes, and the azimuthal order of the mode $m$ such that $m=-\ell, -\ell+1, ..., \ell-1,\ell$, where $\vert{m}\vert$ corresponds to the number of meridian nodes, i.e. showing nodal lines perpendicular to the stellar equator. Radial modes are characterized by $\ell=m=0$ and non-radial modes by $\ell\neq0$. Because of cancellation effects, photometry or radial velocity measurements are not able to access higher degrees of non-radial pulsations with typically  $\ell\geq3$ \citep{kennelly1996}. The only direct way to investigate modes of higher degrees consists in analysing line-profile variations of rotationally-broadened photospheric profiles, where the 3D velocity field of the oscillating stellar surface induces profile variations due to Doppler effect \citep{vogtpenrod1983}. The maximum attainable degree $\ell$ is increasing with the ratio \vsini~ to resolved spectral element ($\Delta v = {\rm{c}}/R$, with $R$ being the resolution of the spectrograph). This also implies that fast rotators seen almost pole-on only allow mode detection limited to low degrees, due to a lack of line profile width.  

\begin{figure*}[!ht]
\centering
 \includegraphics[width=70mm,height=85mm]{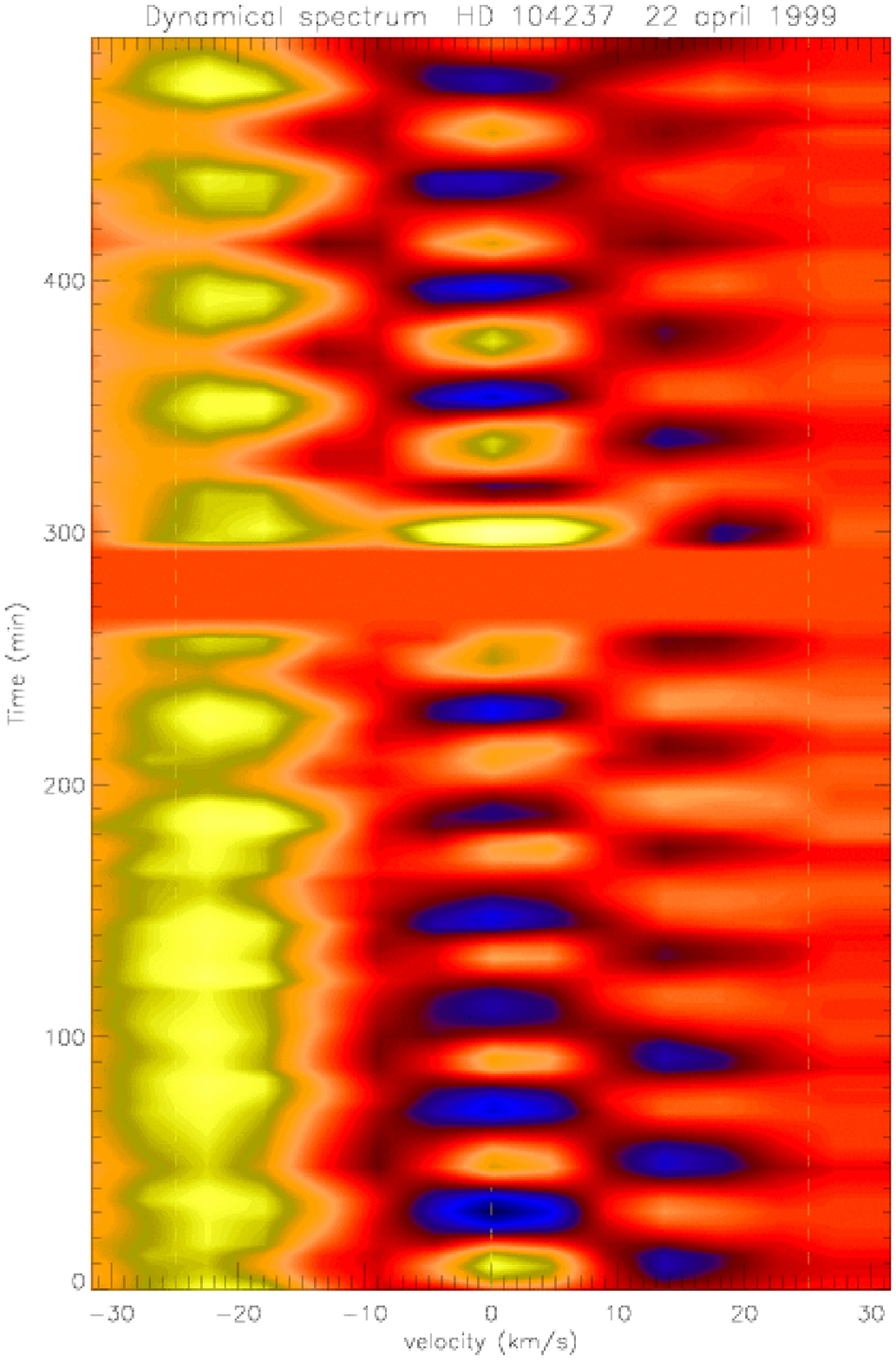}
 \includegraphics[width=70mm,height=100mm]{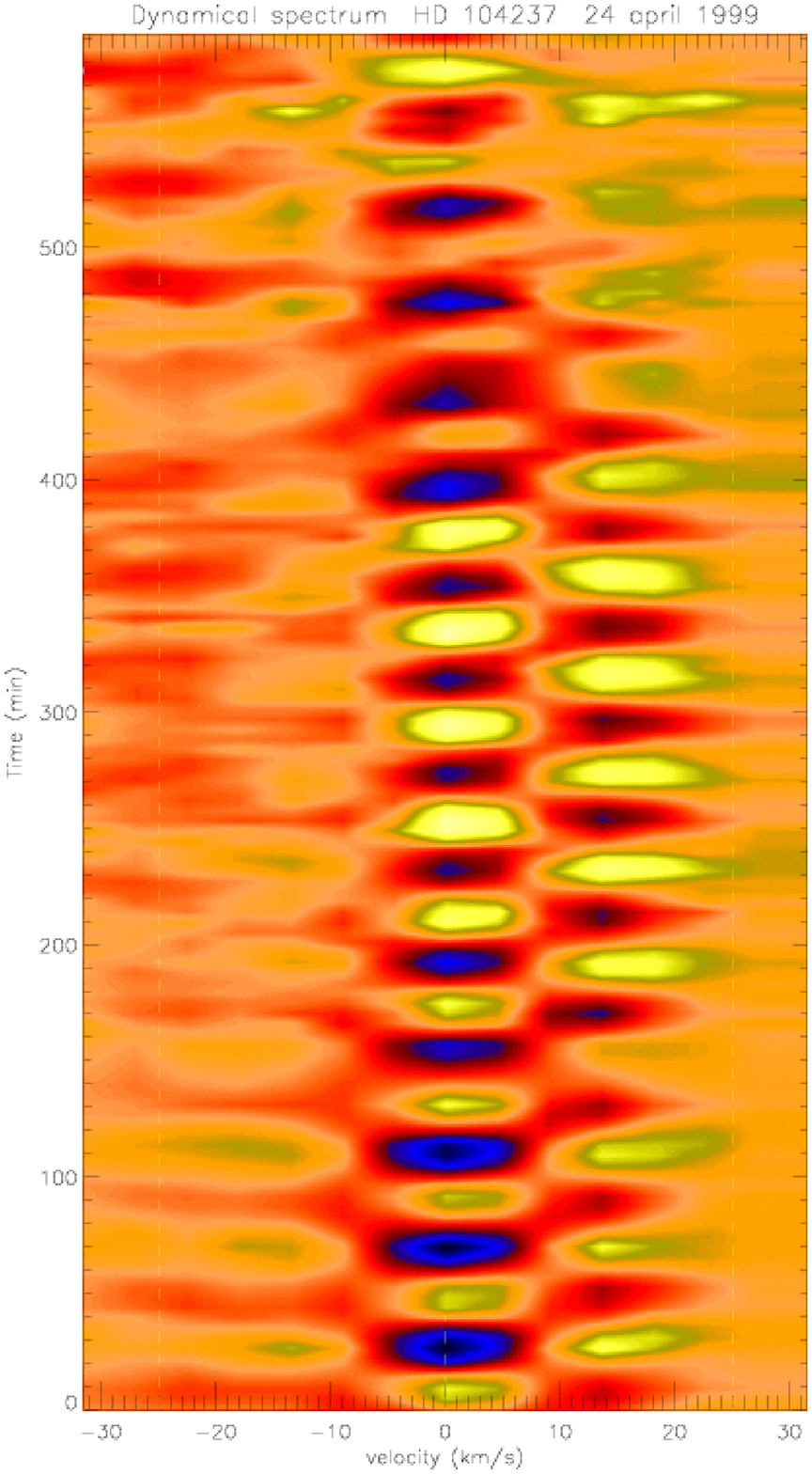}
 \includegraphics[width=70mm,height=103mm]{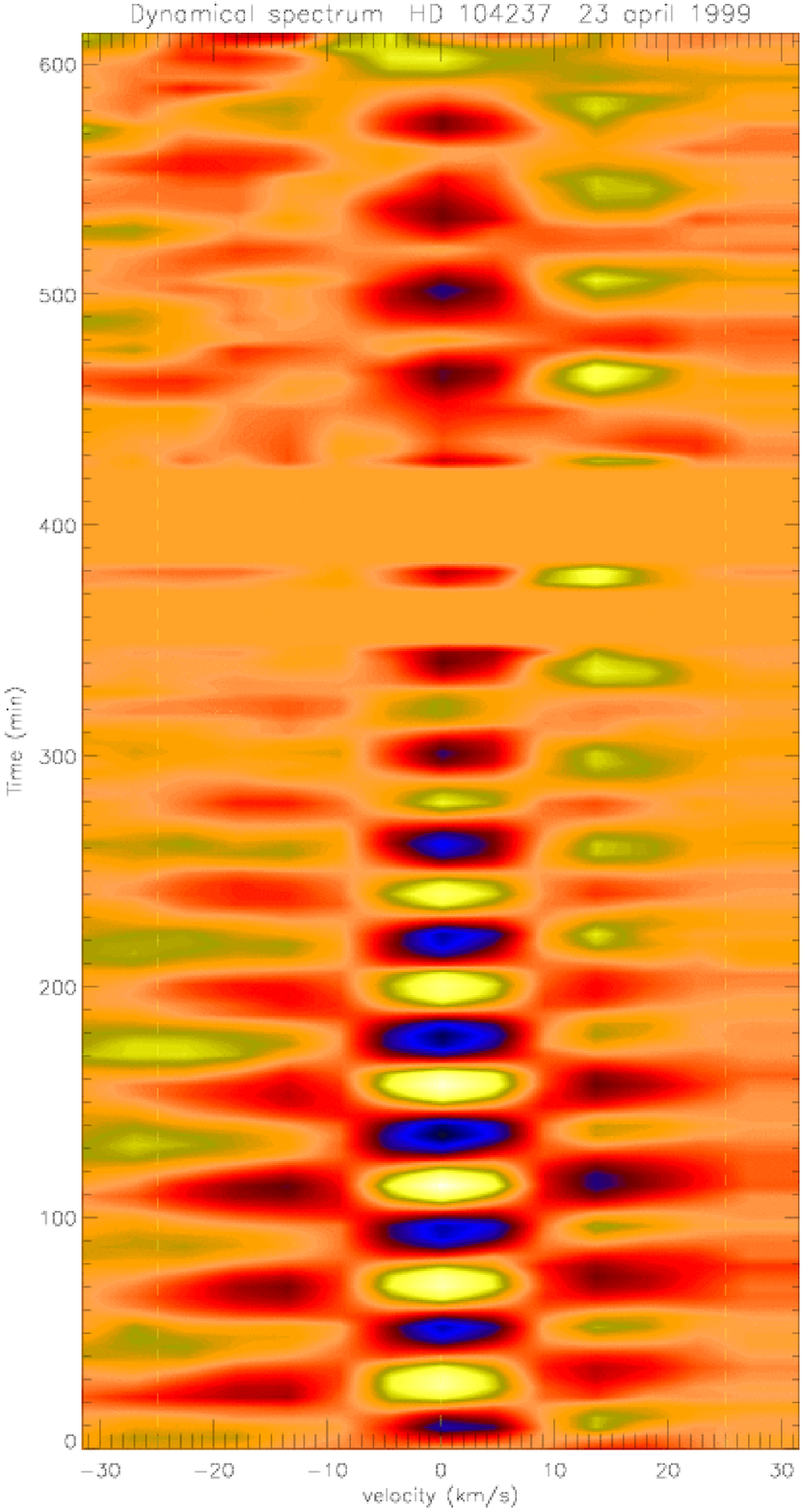}
 \includegraphics[width=70mm,height=110mm]{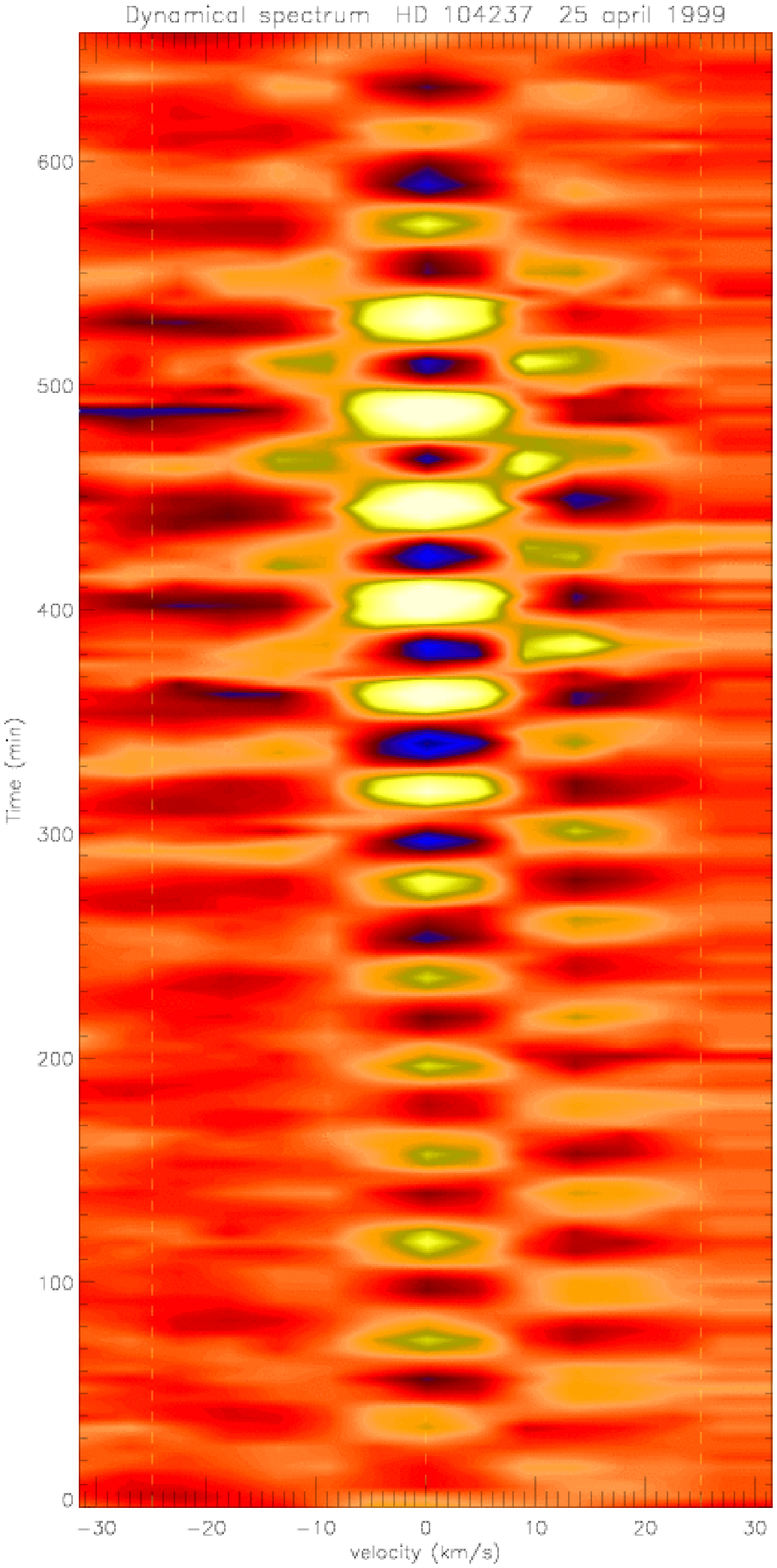}
\caption{Line profile variations due to the non-radial pulsations of HD\,104237. The deviations from the mean intensity are displayed for four different nights (\emph{top left}: April 22$^\mathrm{nd}$ 1999, \emph{bottom left}: 23$^\mathrm{rd}$ 1999, \emph{top right}: April 24$^\mathrm{th}$, \emph{bottom right}: 25$^\mathrm{th}$ 1999).}
\label{lpv2225}
\end{figure*}

\begin{figure}[!h]
 \includegraphics[width=8cm]{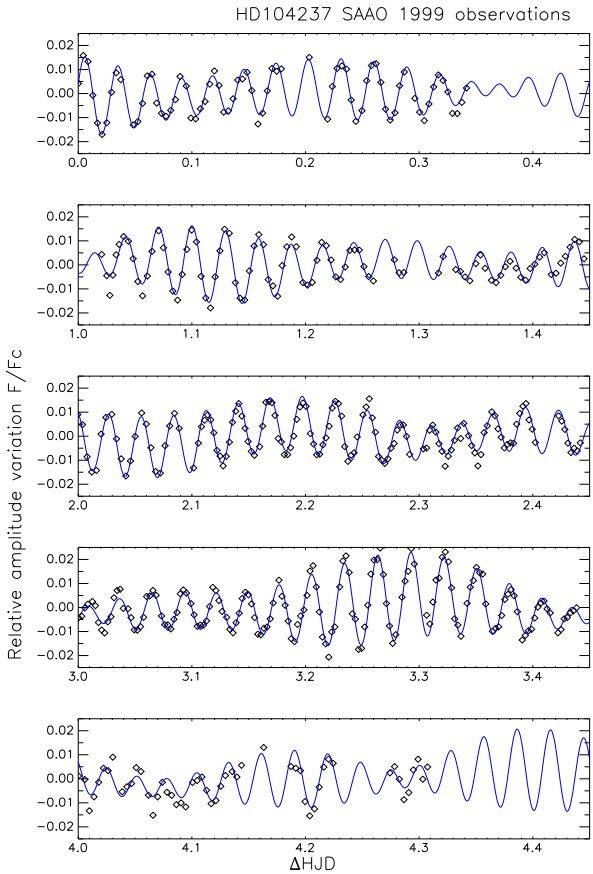}
\caption{Amplitude variation of the central velocity bins of the residual LSD profile of the nights of Apr. $22^{\rm{nd}}$ to  $26^{\rm{th}}$ 1999. Superimposed is a fit of the data corresponding to the result of the frequency analysis as presented in Tab. \ref{tabfreq}. }
\label{ampvar2225}
\end{figure}

As explained in previous Section, we concentrated our study on the data of the $22^{\rm{nd}}$-$26^{\rm{th}}$ of April 1999. The initial work by 
\citealt{bohm2004} did not reveal any non-radial pulsation at first sight in the line profiles and these authors therefore decided to concentrate on radial velocity variations, also due to the fact that the star is a moderate rotator, seen 3/4 pole on and exhibiting a small \vsini~ of only 12$\pm$2 \kms. However, line profiles of this star are broadened in the wings by some additional agent, leading to a total width of almost 50\,\kms, the double of the pure rotational broadening. 
The resolution of the GIRAFFE spectrograph at SAAO being of the order of ${\rm R}\approx35000$, a resolved element corresponds roughly to 8\,\kms. This implies that a search for non radial pulsations (NRP) in this instrumental context is limited in this star to degrees $\ell$ of the order of 2 to 3, higher degrees being not observable due to a lack of rotational velocity resolution.
In the present work we decided nevertheless to go one step further, and to reanalyse in depth the profile variations themselves, not limiting the study to radial velocity variations. To do so, and in order to search for faint traces of non-radial pulsations in the primary component, each LSD profile has been corrected for its radial velocity. This correction comprises orbital motion, but also centroid shifts mainly due to radial pulsations.  After recentering, a mean LSD profile, averaged over the whole run, has been subtracted from every profile in
order to search for very faint signatures. This allowed us to detect line profile variations clearly indicative of non-radial stellar pulsations, inducing differential amplitudes of the order of 1.5\% of the continuum. Also, it is obvious from the differential dynamical spectra of Fig.\ref{lpv2225} that beating between nearby frequencies occur, which leads to the conclusion that probably more than one NRP must be acting (the night of April 26$^{\rm th}$ is not represented due to shorter coverage). The sampling of the LSD profiles in dispersion direction satisfies the Nyquist criteria, i.e. one resolved element projects on 2 velocity bins of 4\kms each. Fig.\ref{ampvar2225} shows the differential amplitude variations at line center, extracted by taking the median value of the 3 central velocity bins of each spectrum. 

\subsection{Central profile analysis with SigSpec and Period04}

We choose to carry out a frequency analysis of our LSD residual profiles time-series using the \emph{Period04} package \citep{lenzbreger2005} in parallel with the \emph{Sigspec} package \citep{reegen2007}. They both use, in iterative process, combination of Discrete Fourier Transform (DFT) and least-squares fitting algorithms to extract the frequencies, amplitudes and phases of multi-periodic signals in not-equally spaced data sets. Both process fit simultaneously all the frequencies, amplitudes and phases detected so far when a new peak is found in the DFT spectrum of the time-serie. Unlike \emph{Period04} which provides a flexible interface to perform multiple-frequency fits, \emph{SigSpec} is fully automatic. In addition, it includes a rigorous statistical treatment of how to compute the significance level of these peaks (with respect to white noise) and provides directly a significance parameters, the spectral significance \emph{sig}.
In order to avoid at maximum low frequency night-to-night variations, we corrected our central velocity bin time series by subtracting a nightly average. Still, by doing so, some low frequencies persisted as can be seen later on. 

\begin{figure}[!h]
 \includegraphics[width=8cm]{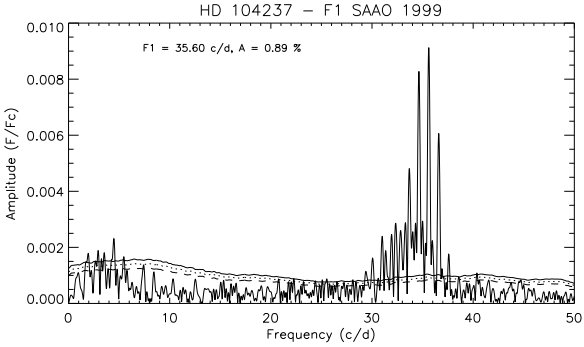}
\caption{Periodogram of the line center variation as shown in Fig.\ref{ampvar2225}, without prewithening. The large bulk of frequencies around 29 - 36 \cd can be seen, as well as the 3.2, 3.6 and 4.0 mean-ampitude level of the noise determined by prewithening F$_{1}$ to F$_{9}$.}
\label{allfreq}
\end{figure}

\begin{figure}[!h]
 \includegraphics[width=8cm]{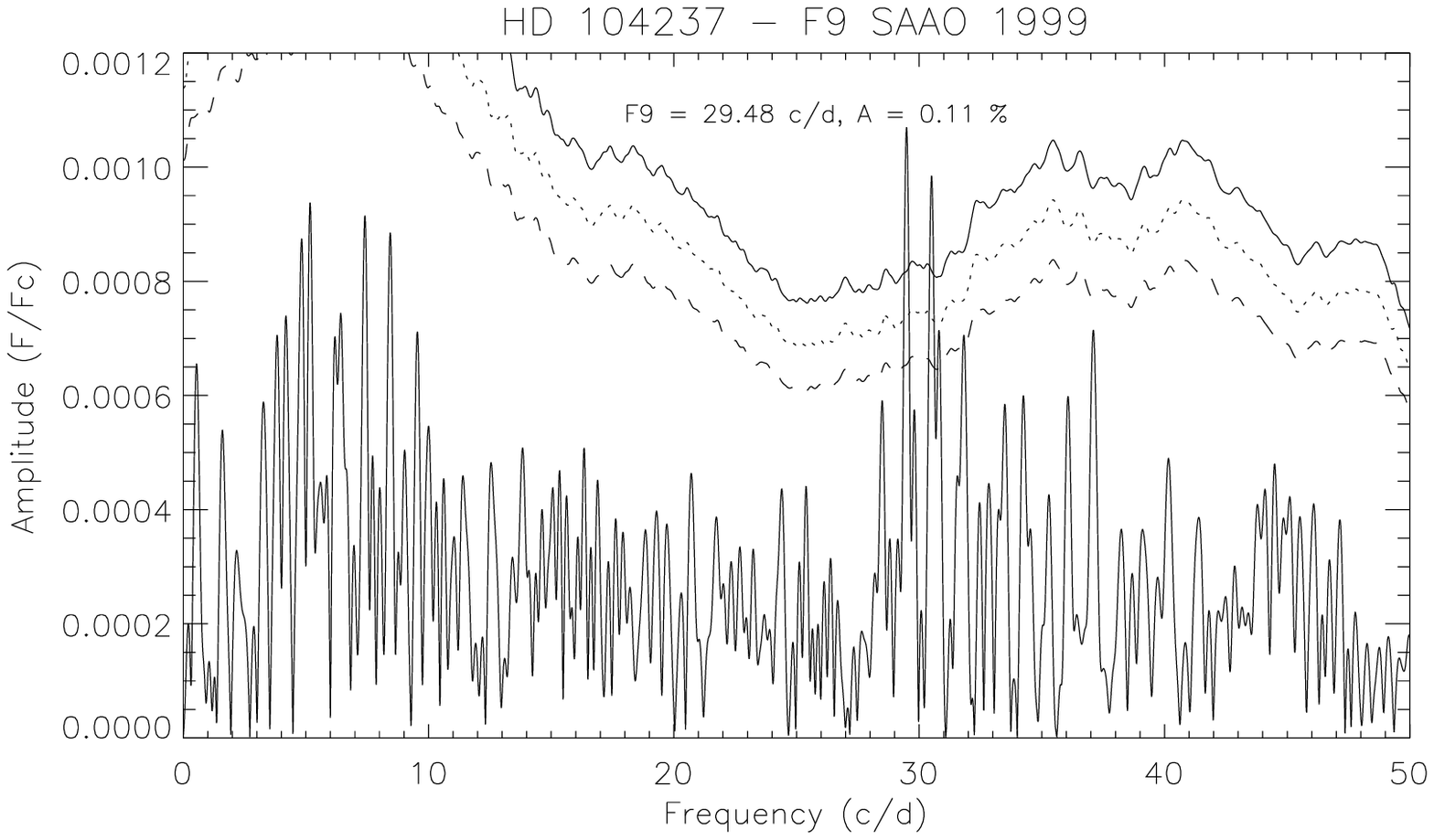}
\caption{Same figure as Fig. \ref{allfreq}, but prewithened by F$_{1}$ to F$_{8}$. Only F$_{9}$ is present, well above the 99.9\% confidence level.}
\label{lastfreq}
\end{figure}

The results of the \emph{Period04} analysis are presented in Table~\ref{tabfreq} based on the amplitude variations as shown in Fig. \ref{ampvar2225}. As described in \citealt{bohm2004}, the uncertainty in the extracted frequency value can be estimated in different ways. The most conservative approach has been proposed by \citep{Ripepi2003} who suggest to estimate the error by measuring the FWHM of the main lobe of the spectral window function, corresponding here to 0.31\,\cd, while \citep{kurtzmuller1999} use as  an estimator the time basis $\Delta T$ of the observing run, yielding for the SAAO 1999 run a much smaller error estimate of 1/(4 $\Delta T$) = 0.04\,\cd. 
\cite{breger1993} studied empirically the possibility of a peak in the periodogram being a true signal of pulsations with respect to the noise level; this work was refined later by \cite{kuschnig}: the significance of a peak in the amplitude periodogram exceeding 4.0 times the mean noise amplitude level after prewhitening of all local frequencies has a 99.9\% probability to be due to stellar pulsations  (99.0\% for a ratio of 3.6, 90.0\% for a ratio of  3.2). Fig. \ref{allfreq} shows the amplitude power spectrum including all strong peaks around 29 - 36 \cd, the noise levels having been determined after prewithening of the 9 first frequencies, while Fig. \ref{lastfreq} shows the fact that even F$_{9}$ has a signal to noise level exceeding 4.0. We arbitrarily decided to stop the iterative procedure for amplitude values below 1\%. The verification of the results with \emph{SigSpec} revealed exactly the same frequencies and amplitudes, all detected frequencies F$_{1}$ to F$_{10}$ having a significance level between 66.5 and 6.3, respectively.  No combined frequency is present in the power spectrum, following the analysis with the  \emph{Combine} program by  \cite{reegen2007}. We don't attribute much importance to the frequencies below approximately 5\,\cd, since differential calibration issues from one night to the other might persist, i.e. this excludes F$_{4}$, F$_{7}$, F$_{8}$ and F$_{10}$. By applying the more conservative error estimate by \citep{Ripepi2003}, all remaining frequencies ("F") could be identified with the frequencies ("f") determined by radial velocity studies on the same data set in \cite{bohm2004}, F$_{3}$ being slightly outside the error bar, and the 1.0\,\cd difference between F$_{9}$ and f$_{3}$ indicates that we are most likely in presence of an alias of the same frequency. It can be noticed, that the order of the first two dominant frequencies is inverted depending if radial velocity or central depth variations are analysed.\\

\begin{table}[!h]
\centering
\caption{Frequencies determined with \emph{Period04}. The different columns are: (1) number, (2) frequency in \cd and (3) in $\mu$Hz. (4) Amplitude in  (F/F$_{\rm{c}}$) (flux variation with respect to the average profile), (5) identification with frequencies as of \cite{bohm2004}, and absolute shift in \cd.} \label{tabfreq}
\begin{tabular}{ccccc}
\hline\hline
&&&&\\
 \#   &F$_{\rm P04}$  & F$_{\rm P04}$ & A$_{\rm P04}$      & f$_{\rm{B04}}$ ($|\delta \rm{f}|$)\\
&(\cd)              &($\mu$Hz)&(F/F$_{\rm{c}}$) &     \\ 
&&&&\\ \hline
&&&&\\
F$_{1}$ 	& 35.60         &  412.04  &0.0089 &                    f$_{2}$ (0.01)\\
F$_{2}$ 	& 33.74         &   390.51 &0.0033 &                    f$_{1}$ (0.12)\\
F$_{3}$ 	& 32.25         &   373.26 &0.0030&                     f$_{6}$ (0.36)\\
F$_{4}$ 	& 4.47           &    51.74  &0.0024 &                     \\
F$_{5}$ 	&31.08         &   359.72 &0.0028 &                   f$_{4}$ (0.12)\\
F$_{6}$ 	& 34.00         &   393.52 &0.0019&                    f$_{5}$ (0.14)\\
F$_{7}$ 	& 1.94           &     22.45  &0.0018 &                   \\
F$_{8}$ 	& 2.72           &     31.48  &0.0014 &                   \\
F$_{9}$ 	& 29.48         &    341.20 &0.0011&                   f$_{3}$ (1.0)\\
F$_{10}$ 	& 5.17           &       59.84 &0.0010&                   \\
&&&&\\
\hline
\end{tabular}
\end{table}

\subsection{Mode identification with the Fourier 2D method}

\begin{figure}[!h]
\includegraphics[width=80mm,height=80mm]{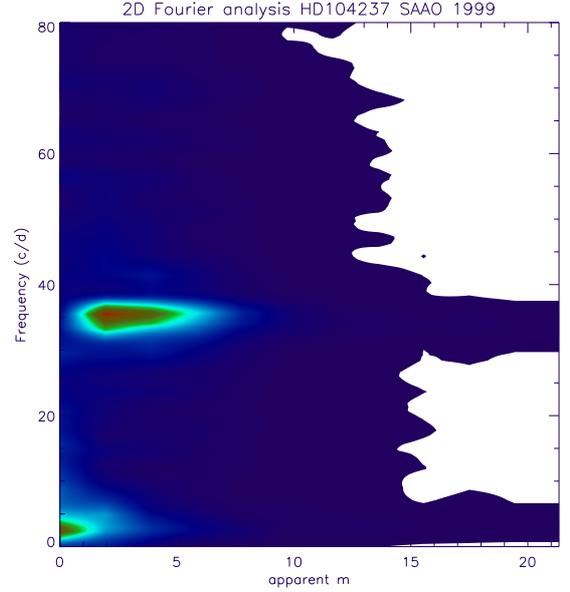}
\caption{Combined F2D analysis of HD\,104237 for the nights of Apr. $22^{\rm{nd}}$ to  $25^{\rm{th}}$ 1999. }
\label{fouall}
\end{figure}

A direct method of analysing the non-radial pulsation modes present in the LSD-spectra time series consists in applying the Fourier 2D method on line-profile variations \citep{kennelly1993,kennelly1994,kennelly1996}. This technique analyses the complex pattern present in the line profiles by computing a 2-dimensional Fourier transform in both time and Doppler space. To do this, an interpolation of each profile on a grid representing stellar longitudes is performed, transforming velocities across the line profile into longitudes on the stellar equator using the relation  $\Delta v = v\sin i  \sin\phi$. $\Delta v$ is here the velocity position within the LSD profile with respect to the rest wavelength of the star and $\phi$ is the stellar longitude angle of the star in spherical coordinates. In the resulting two-dimensional Fourier spectrum,  the temporal frequencies are related here mainly to the frequencies of oscillation, while the apparent azimuthal order \^m is related to the structure of the modes present at the stellar surface, without being identical to the usual azimuthal order m. The original work by \citet{kennelly1994} showed in fact that apparent  $\left | m \right |$ scales as $\ell$ +2 for values close to zero, as $\ell$ +1 for values lower than 10 and as $\ell$ for values above 10. We performed this F2D computation on the night of April $22^{\rm{nd}}$ to $25^{\rm{th}}$ 1999. A weighted combination of the nightly F2D spectra can be seen in Fig.~\ref{fouall} (see \citealt{bohm2009} for details of the method) and our results confirm the presence of low-degree non-radial pulsations: a peak at the dominant frequency F$_{1}$ is clearly seen with an apparent \emph{\^{m}} comprised between 2 and 4 indicating a degree $\ell$ comprised between 0 and 2. Moreover, as can be seen e.g. in Tab. 2.3 of \cite{kennelly1994} (or in simulation with available online-NRP simulators), the symmetric pattern of the dynamical residual profile as seen in Fig.\ref{lpv2225} indicates for F$_{1}$ an azimuthal order m = $\pm$ 1, implying a most likely $\ell$ value of 1 or 2. 

The frequency peak in Fig. ~\ref{fouall} between 3 and 5\,\cd is again not taken into account, since nightly calibration shifts can easily introduce such variations.   

A complementary analysis on the whole data set using the FAMIAS code \citep{zima2006,zima2008} is planned, but the highly asymmetric mean photospheric profile seen in HD\,104237 adds complexity in the data analysis and cannot in the actual state of the code be taken into account.

\section{Optimized spectral continuum determination}
\label{norm}

In addition to the best available mode identification, a future modeling of the stellar oscillations of HD\,104237 requires the precise redetermination of its fundamental stellar parameters. Since we decided to carry out this analysis from reliably measured equivalent width of selected absorption lines (see Sect.~\ref{funda}), a perfect continuum normalization revealed to be essential, particularly in order to significantly increase the SNR by summing all the spectra of the night. An automatic code was therefore developed, the case of HD\,104237 being rather complicated due to its spectroscopic binarity: the pollution of the primary (P) spectrum by the secondary (S), but also the presence of emission components and variability observed in numerous absorption lines require particular attention in the normalization procedure. The data set obtained at SAAO in 2000 (see \citealt{bohm2004}) contained one night of observations very close to the periastron of the binary (12$^{\rm th}$ of April). We decided therefore to concentrate on the data of this particular night in which both spectra are well separated in velocity, making easier to distinguish P from S. As we worked on echelle spectra, we carried out an individual normalization of each order, performed in 4 mainly automated main steps.

As a first step, we created a comb including effective continuum locations and eliminating unsuited others areas. To do so, we extracted 2 lists of absorption lines ($T_{\rm{eff}}$ (P) = 8500\,K and $T_{\rm{eff}}$(S) = 4750\,K, log\,$g$ = 4.0, $v_{\rm{micro}}$ = 2\,\kms and
solar abundances) from the Vienna Atomic Line Database (VALD; \citealt{piskunov1995,ryabchikova1997,kupka1999,kupka2000}) on which we applied:\\
i) a selective criterion on the maximum depth accepted for spectral lines to be considered as continuum ($\leq1$\% of the continuum),
taking into account the relative luminosities of both components as a function of wavelengths, assuming two blackbodies with temperatures equal to 8500 K and 4750 K,\\
ii) a criterion on the atypically large profile width of $\pm 2.5\,v\sin{i}$ excluding areas with spectral lines,\\
iii) the elimination of specific areas whith emission, variable, telluric and the broad Balmer lines (the latter extending over several spectral orders and therefore not reliably to be normalized),\\
iv) a shift of P and S wavelength grids by respective orbital velocities at the given observation time (all orbital parameters of the spectroscopic binary have been determined in \citealt{bohm2004}).

After multiplication of the resulting comb with the observed spectrum, only wavelengths ranges corresponding to the remaining continuum areas were kept, all other areas were rejected. The next step consisted in fitting a polynomial of degree 4 through the remaining points in each order and a division of each order of the observed spectrum by this optimized polynomial. Finally, in a last step the software procedes to a weighted concatenation of the normalized orders, the overlap between orders i and i+1 were taken into account through the following equation :
\begin{equation}
  \rm{F_{overlap}(\lambda)=\frac{F_{i}(\lambda)\times
SNR_{i}^{2}(\lambda) + F_{i+1}(\lambda)\times
SNR_{i+1}^{2}(\lambda)}{SNR_{i}^{2}(\lambda) + SNR_{i+1}^{2}(\lambda)}}
\end{equation}
where $\rm{F_{overlap}}(\lambda)$ is the total flux resulting from the
addition of two overlapped parts of consecutive orders with normalized
flux $\rm{F (\lambda)}$, weighted by their corresponding local SNR($\lambda$), resulting from a polynomial fit of the respective SNR curve within the overlap, the local SNR being part of the reduced spectra files.  

\begin{figure}[!t]
\includegraphics[width=90mm,height=70mm]{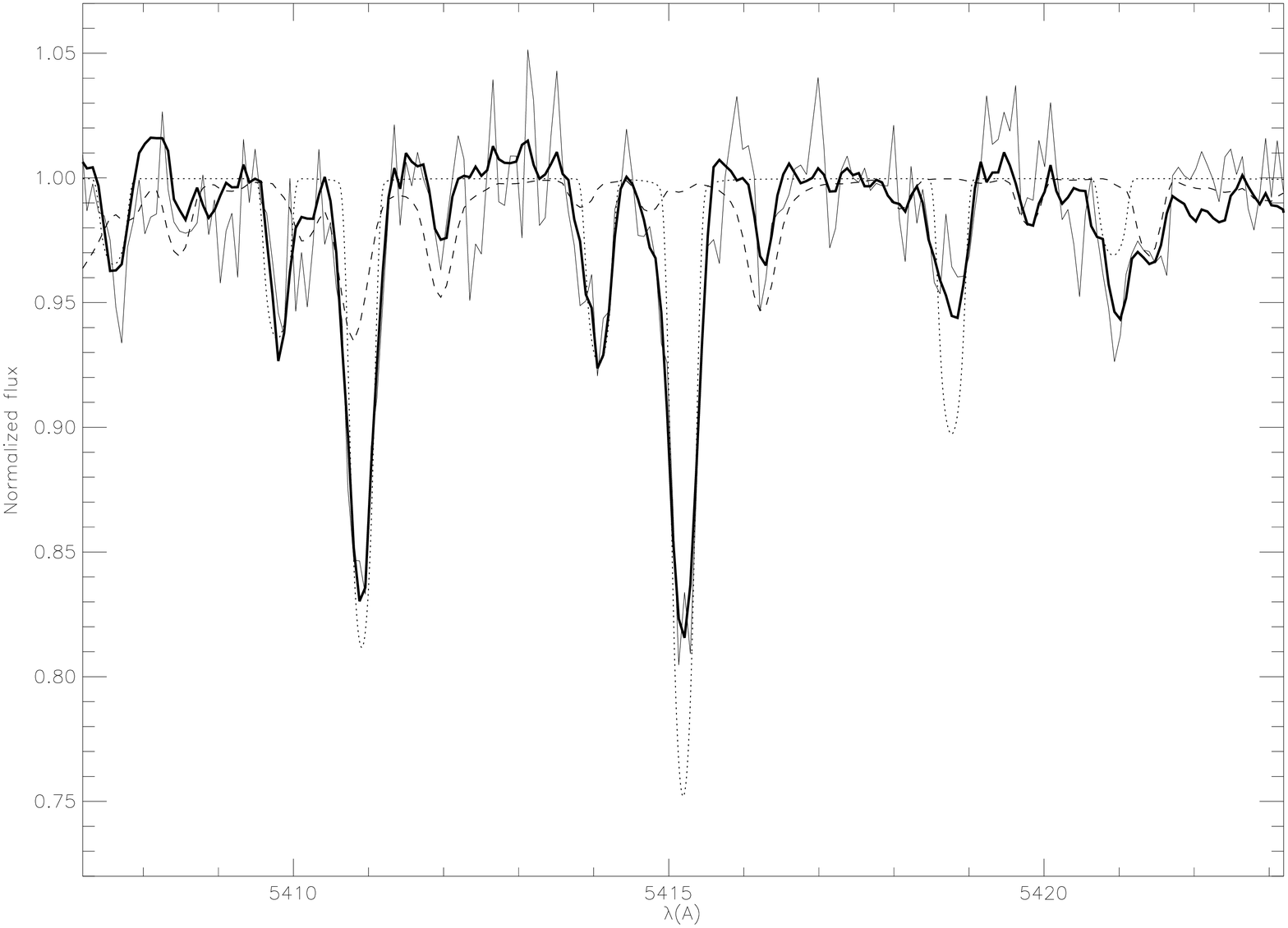}
\caption{Individual (thin line) and summed (thick line) normalized spectrum of the periastron night around one of the selected lines for our study (Fe\,I 5415.1920\,\AA). Synthetic spectra with $T_{\rm{eff}}=$8500\,K (dotted line) and $T_{\rm{eff}}=$4750\,K (dashed line) have been overplotted (\logg =4.0, $v_{\rm{micro}}$ = 2\,\kms solar abundances for both spectra), showing the location of the individual lines. The secondary spectrum has been corrected by the local luminosity ratio. The figure shows that the line selection process described in Sect. \ref{funda} works well. However, it can be noticed that either fundamental parameters and/or local luminosity ratios are not yet perfectly tuned at this stage.
}
\label{figSNR}
\end{figure}

Since Balmer line profiles span several orders, they were not taken into account in this normalization process, i.e. area from 4816 to 4929 \AA\  and from 6475 to 6645 \AA\ remained unnormalized. During periastron both binary components show significant radial velocity shifts within one night. In order to sum up the large number of individual spectra during that night, we decided to center each individual spectrum to the rest wavelength of the primary component. Summation was thereafter done in the following way:
\begin{equation}
  \rm{F_{tot}(\lambda)}=\frac{\sum^{n}_{j=1}F_{j}(\lambda)\times
SNR_{j}^{2}}{\sum^{n}_{j=1}SNR_{j}^{2}}
\end{equation}
where F$_{\rm{tot}}(\lambda)$ is, at each wavelength, the summed normalized spectrum, n is the number of spectra
of the night, $\rm{F (\lambda)}$ the individual flux, SNR is the mean SNR of each spectrum.
After summation, the gain in SNR of our spectrum was of $\approx 6$ since the SNR per resolved element at 550 nm increased from 64 to 372 (from 45 to 263 per pixel; see Fig.~\ref{figSNR}).

\section{Fundamental parameters determination}

\label{funda}

An accurate and reliable knowledge of stellar fundamental parameters, mainly effective temperature \teff, surface gravity \logg~ and chemical abundances $\log A$ (defined as follows: $\log A_{{\rm{E}}}=\log{\left( \frac{N_{{\rm{E}}}}{N_{{\rm{H}}}} \right)}$ for an element E) is crucial to constrain the stellar atmosphere model, which is the basic ingredient of a forthcoming asteroseismic modeling. As shown in Sect.~\ref{previous},  for HD\,104237 large discrepancies persist between parameters announced in the literature. A precise redetermination of its fundamental parameters is therefore crucial. 

\subsection{Method}

\begin{table*}[!t]
\centering
\caption{Characteristics of absorption lines selected for fundamental parameters determination. Column are: (1) central wavelengh, (2)  ion, (3) excitation potential (eV), (4) log of statistical weight $\times$ oscillator strength of the line transition, (5) local mean Signal to Noise, (6) observed equivalent width, (7) absolute and (8) relative uncertainty on the equivalent width.} \label{tablelines}
\begin{tabular}{c|l|c|c|c|c|c|c}
\hline
$\lambda_{\rm{c}}$ (\AA) & el.$+$ion. & $\chi_{\rm{eV}}$ & $\log gf$ & $\overline{SNR}$ & EW$_{\rm{obs}}$ (m\AA) & $\sigma_{\rm{abs}} \rm{(EW_{obs})}$ (m\AA) & $\sigma_{\rm{rel}} \rm{(EW_{obs})}$ (\%) \\
\hline \hline
 4485.675 & Fe\,I  & 3.686 & -1.020 & 217 & 16.1 &  2.6 & 15.9\\ \hline
 4602.941 & Fe\,I  & 1.485 & -2.209 & 245 & 39.3 &  4.0 & 10.1\\ \hline
 5049.819 & Fe\,I  & 2.279 & -1.355 & 268 & 51.9 &  3.9 &   7.6\\ \hline
 5132.669 & Fe\,II & 2.807 & -4.094 & 275 & 28.0 &  4.0 & 14.5 \\ \hline
 5393.167 & Fe\,I  & 3.241 & -0.715 & 289 & 46.7 &  3.1 & 6.6 \\ \hline
 5415.192 & Fe\,I  & 4.386 &  0.642 & 285 & 97.9 &  4.4 & 4.5 \\ \hline
 5434.523 & Fe\,I  & 1.011 & -2.122 & 273 & 51.9 &  3.3 & 6.3 \\ \hline
 5576.089 & Fe\,I  & 3.430 & -1.000 & 272 & 35.2 & 3.5 & 9.8 \\ \hline
 6084.111 & Fe\,II & 3.199 & -3.881 & 245 & 18.2 &  4.0 & 21.8 \\ \hline
 6400.000 & Fe\,I  & 3.602 & -0.290 & 238 & 51.3 &  4.7 & 9.2 \\
\hline
\end{tabular}
\end{table*}

Since \teff, \logg, chemical abundances ($\log A$) and microturbulent velocity $v_{\rm{micro}}$ are interdependant and affect simultaneously the        
depth, width and shape of spectral absorption lines, they have therefore to be determined simultaneously. Numerous methods exist for such an analysis, but very few of them are suitable in the particular case of HD\,104237.
Due to the non standard shape of HAeBe spectral energy distributions (IR excesses, UV depletion), classical photometric determination of \teff\ have to be considered with caution. Another frequently used method, the stellar parameter determination using Balmer lines have to be excluded because of core emission in the line and/or P Cygni profiles, but also on a technical side due to normalization issues in case of high dispersion echelle spectra. Techniques based on lines ratio for determining \teff\ (see e.g. \citealt{sousa2010}) require non polluted lines and were rejected because of strict selection criteria (see Subsect.~\ref{sel}), which excluded all workable lines in our case. 

After consideration, we favoured an equivalent width (EW) analysis with respect to a direct spectrum fitting method, because of the existence of additional broadening agents and strong asymmetries in the line profiles, whose origin is unknown and which are therefore difficult to reproduce, and of numerous activity features in the whole spectrum that  complicate direct line profile fitting. Our method consist in a comparison of EW of a set of selected lines, between the observed combined spectrum of the periastron night of HD\,104237 and a fine grid of synthetic spectra in the 3 fundamental parameter dimensions, namely  \teff,  \logg\ and chemical abundance of a given chemical element (required to work out reliable values of \teff\ and \logg). With more than 2000 spectral lines featured in a 8500\,K synthetic spectra, Fe turned out to be the most suitable chemical element to establish our EW analysis. The high activity level of this HAe star (emission, variability, line asymmetry) and its binarity make a reliable determination of fundamental parameters in this case very challenging.

\subsubsection{Correction of the secondary spectral contribution}
Since our main goal was to determine the fundamental parameters of the primary component, we had to correct for the polluting spectroscopic contribution of the secondary component, HD\,104237b. Indeed, although the secondary is significantly fainter than the primary component, it is also much cooler and does exhibit strong metallic absorption lines, leading to systematic contamination of primary lines and even lowering in some parts of the spectrum the continuum level. To free the observed spectrum from this effect, a synthetic spectrum corresponding to the secondary component was substracted from the observed spectrum. In order to minimise additional errors from an incorrect model of secondary, we used stellar parameters determined by \cite{bohm2004} based on cinematics of the binary movement, adopting solar chemical abundances, a \logg\ value of 4.0 and a $v_{\rm{micro}}$ of 2 \kms: several synthetic spectra, with \teff(S) comprised between 4250\,K and 5250\,K (by step of 250\,K) were therefore tested. Since the observed normalized spectra had been centered on the rest wavelength of the primary component, secondary synthetic spectra were shifted by the respective orbital velocity displacement. 
We called  $\alpha(\lambda)$ a monochromatic luminosity ratio with $\alpha(\lambda) = B_{\rm{P}}(\lambda)/B_{\rm{S}}(\lambda)$, where 
$B_{\rm{P}}(\lambda)$ and $B_{\rm{S}}(\lambda)$ are respectively the Planck laws of the primary and the secondary. The most likely secondary model was determined by computing and minimizing a $\chi^{2}$ quantity between the observed spectrum and the expected secondary contribution, yielding a \teff\ value of 4500\,K for the secondary. A spectrum corresponding to this temperature ($\rm{F_{S}(\lambda)}$ flux) was therefore subtracted from the observed spectra ($\rm{F_{obs}(\lambda)}$ flux), with a contribution based on the equation:
\begin{equation}
    \rm{ F_{obs}(\lambda)=\frac{F_{P}(\lambda)+F_{S}(\lambda)/\alpha(\lambda)}{1+1/\alpha(\lambda)} }
\end{equation}
in order to obtain a primary spectrum ($\rm{F_{P}(\lambda)}$ flux) cleaned up from the secondary contribution. Obviously, systematic uncertainties of the monochromatic luminosity ratio and  the intrinsic stellar parameter selection of the secondary do persist and are difficult to fully be taken into account.

\subsubsection{Construction of a 3D-grid of synthetic spectra}

In order to precisely redetermine the fundamental parameters of our star we needed to compare the observed high SNR spectrum to synthetic spectra.
Therefore, we constructed a 3-dimensional grid (\teff, \logg, $\log{A}_{\rm{Fe}}$) of synthetic spectra based on line catalogs provided by the VALD database (see Sect.\ref{norm}) as input of the SYNTH3 spectrum synthesis code \citep{kochukhov2007}, each spectrum covering a wavelength range from 4400\,\AA\ to 7000\,\AA. The SYNTH3 code assumes Local Thermodynamic Equilibrium (LTE) and a plane-parallel hydrostatic stellar model atmosphere. For early type stars input models are provided by the well known Kurucz program suite. We selected solar abundance atmosphere models from the grid, after having checked that the effect of an input model containing slightly different than solar metallicities is of second order on the calculated synthetic spectrum.  

As a starting point, we adopted parameters derived by previous authors to define the boundaries of the grid: 7500\,K\,$\leq$\,\teff\,$\leq$\,9500\,K (by steps of 250\,K) and 3.5\,$\leq$\,\logg\,$\leq$\,4.5 (by steps of 0.5). Since it is not uncommon to find HAeBe stars exhibiting in their spectra simultaneously the presence of some elements in suprasolar  abundance, whereas other ones are subsolar (see e.g. \citealt{ackewaelkens2004}), without particular trend in terms of global metallicity, we decided to modify in this study only the Fe abundance in our VALD requests. It should be noticed at this stage, that studying the iron abundancy was a necessary by-product in order to constrain effective temperature, as we will see subsequently. 
$\log{A}_{\rm{Fe}}$ values comprised between $-4.84$ and $-4.00$ seemed to be reasonable, i.e. [Fe/H] values comprised between -0.3\, and +0.54\, with respect to the solar Fe abundance given in VALD, which adopted $\log{A}_{\rm{Fe_\odot}}$ = -4.54. As a reminder, [Fe/H]=$\log{\left(\frac{N_{{\rm{Fe}}}}{N_{{\rm{H}}}}\right)_{\star}}-\log{\left(\frac{N_{{\rm{Fe}}}}{N_{{\rm{H}}}}\right)_{\odot}}$.
Typical values of $v_{\rm{micro}}$ found in the literature were of 2-3 \kms (see e.g. \citealt{ackewaelkens2004,guimaraes2006,catala2007}), we therefore fixed a value of 2 \kms for the microturbulent velocity HD\,104237.
A wavelength by wavelength quadratic interpolation was subsequently done between synthetic spectra in order to obtain a finer spectral grid with a 25\,K step in \teff, a 0.1 step in \logg\ and a 0.01 step in $\log{A}_{\rm{Fe}}$.

\subsubsection{Photospheric line selection}
\label{sel}
Photospheric iron absorption lines used in our study were selected using the following criteria:
\begin{itemize}
\item[--] exist for every (\teff, \logg, $\log \rm{A_{Fe}}$) triplet in order to be intercomparable;
\item[--] exclude blends with lines from other chemical elements (than Fe), i.e. within a range of $\pm 2.3\, v\sin i/\rm{c}$ from the central wavelength (the extraction width having been adoped after visually inspecting the spectra and taking into account the broad wings of the photospheric lines of our star);
\item[--] avoid pollution by lines from the secondary spectrum, even if a best-fit secondary spectrum had been removed, in order to minimise the
impact of the potential error on \teff(S). To do so, secondary lines with a relative contribution deeper than 10\% of the central depth led to a rejection of the primary spectrum line.
\item[--] show no or little variability in time (especially over short periods of time);
\item[--] do not exhibit at any time an emission or other than  normal (for HD\,104237) asymmetry; 
\item[--] be strong enough with respect to local noise, in order to minimize relative uncertainties on EW and to be able to neglect residual contribution from very faint lines of primary and secondary spectra (VALD requests require a depth threshold and provide catalog of lines deeper than this threshold value), but not to strong to avoid saturation effects. As a result, we kept weak and moderratly strong lines (15\,m\AA\,$\leq$\,EW$_{\rm{obs}}\leq$\,100\,m\AA);
\item[--] show exploitable nearby continuum, e.g. with no emission component in their immediate vinicity (after automatic selection, a individual visual check line by line was done);
\item[--] cover a variety of excitation potentials, ionization stages and wavelengths;
\item[--] having ascertained atomic parameters, in particular $\log gf$ values (comparison with \citealt{fuhrwiese2006} and \citealt{melendezbarbuy2009}).
\end{itemize}
This rigorous selection resulted in a set of only 10 spectral lines (8 Fe\ I lines and 2 Fe\ II lines) which are listed in Table~\ref{tablelines}.


\subsubsection{EW measurements}
\label{EWmeasur}

EW measurements of our selected lines were carried out using a trapezoidal integration. For consistency, the identical extraction limits were used for the integration of a given line from both the observed spectrum and the synthetic spectra.

To extract errors on EW values, we assumed that they mainly came from local averaged photon noise and we neglected uncertainties in integration limits or arising from residual continuum normalization errors. Multiplying the local $\overline{\rm SNR}$ value by the line integration width $\Delta v$ of each line, we then obtained an estimation of absolute uncertainty in EW, \,$\rm{\sigma_{abs}(EW_{obs})}$, and its corresponding relative uncertainty $\rm{\sigma_{rel}(EW_{obs})=\sigma_{abs}(EW_{obs})/EW_{obs}}$ (see Table~\ref{tablelines}).

\subsection{\teff, \logg\ and Fe abundance determination}
\label{determin}
In order to determine \teff, \logg\ and log A$_{\rm{Fe}}$ of HD\,104237, it was necessary to compute the degree of similarity between our observed spectrum and all the synthetic spectra of our 3D-grid by means of the 10 previously measured EW;  a direct synthetic spectrum fitting technique could not be applied due to highly atypical photospheric line profiles (asymmetry, broad wings). 

A maximum likelihood estimation was therefore done by minimizing the following reduced merit function $S_{\rm{red}}$:
\begin{equation}
   S_{\rm{red}}=\frac{1}{{\rm{N-k}}}\sum^{{\rm{N}}}_{{\rm{j}}=1} \left( \frac{{\rm{EW}}_{{\rm{obs,\,j}}}-{\rm{EW}}_{{\rm{synth,\,j}}}(T_{\rm{eff}}, \rm{log}\,g, \rm{log\,A_{Fe}})}{\sigma_{j}} \right)^{2} 
\end{equation}
where N-k is the number of degrees of freedom, N being the number of selected lines and k being the number of parameters to be determined (3 in our case).
A good introduction to this statistical approach can be found in \cite{press1992}.
EW$_{\rm{obs}}$ and EW$_{\rm{synth}}$ are respectively the EW of the observed spectrum and the EW of a synthetic spectrum, $S_{\rm{red}}$ being computed for each synthetic spectrum of our 3D-grid. Lastly, $\sigma$ corresponds to the uncertainty on both EW$_{\rm{obs}}$ and EW$_{\rm{synth}}$ such that $\sigma^{2}=\sigma_{{\rm{obs}}}^{2}+\sigma_{{\rm{synth}}}^{2}$. Because of the optimal normalization and the absence of noise in synthetic spectra, $\sigma_{{\rm{synth}}}$ was neglected and we assumed that $\sigma=\sigma_{{\rm{obs}}}$, $\sigma_{{\rm{obs}}}$ being described in Sect.~\ref{EWmeasur} and in Table~\ref{tablelines} as $\rm{\sigma_{abs}(EW_{obs})}$.
The error bars on the parameters of the synthetic spectrum yielding the lowest value of $S_{\rm{red}}$ were then investigated by means of statistical tools. The merit function $S_{\rm{red}}$ follows a $\chi^{2}$ distribution with $\nu=$N-k degrees of freedom if: i) we consider that the best model supplied by the minimum merit function is the right physical model representing the star and therefore that the corresponding stellar parameters are the true parameters of the star, ii) we assume that the perturbations on ($\rm{EW_{obs}-EW_{synth}}$) are independent and additive standard normal random variables. Since the $\sigma_{\rm{obs}}$ values are estimated from the local SNR around the selected lines and since the noise is poissonian and tends to a gaussian distribution in case of a great number of photons, which is the case here, the gaussian assumption is valid in this study. In this situation, the error bar corresponding to a choosen confidence interval $x$ can be provided by the cumulative distribution function of the $\chi^{2}$ distribution with $\nu$ degrees of freedom defined as:
\begin{equation}
   D_{\nu}(x)=\int_{0}^{x}\frac{t^{\frac{\nu}{2}-1}e^{-\frac{t}{2}}}{\Gamma(\frac{\nu}{2})2^{\frac{\nu}{2}}}dt=\frac{\gamma(\frac{\nu}{2},\frac{x}{2})}{\Gamma(\frac{\nu}{2})}
\end{equation}
where $\Gamma$ and $\gamma$ denote the Gamma and lower incomplete Gamma function, respectively. $D_{\nu}(x)$ is equal to the probabilty $\mathbb{P}(S \leq x)$ of the merit function $S=S_{\rm{red}}\times ({\rm{N-k}})$ to be within the confidence interval $x$. By misuse of langage, such confidence intervals on the $S$ value are usually considered to be equivalent to confidence intervals on the resulting parameters. For the needs of our study, we did the same assumption. We were therefore searching for the interval $\Delta S$ such that $S_{\rm{min}}+\Delta S \leq x$, which was equivalent to search for $\Delta S_{\rm{red}}=\Delta S/(\rm{N-k})$ such that $S_{\rm{red,\,min}}+\Delta S_{\rm{red}} \leq x/(\rm{N-k})$. With $\nu=10-3=7$ degrees of freedom we deduced that parameters of synthetic spectra included within an interval $\Delta S_{\rm{red}}$ of 1.17, 2.04 and 3.12 from the minimum $S_{\rm{red,\ min}}$ value were respectively 68.3\%, 95.4\% and 99.7\% reliable.

\begin{figure*}
\centering
\includegraphics[width=150mm]{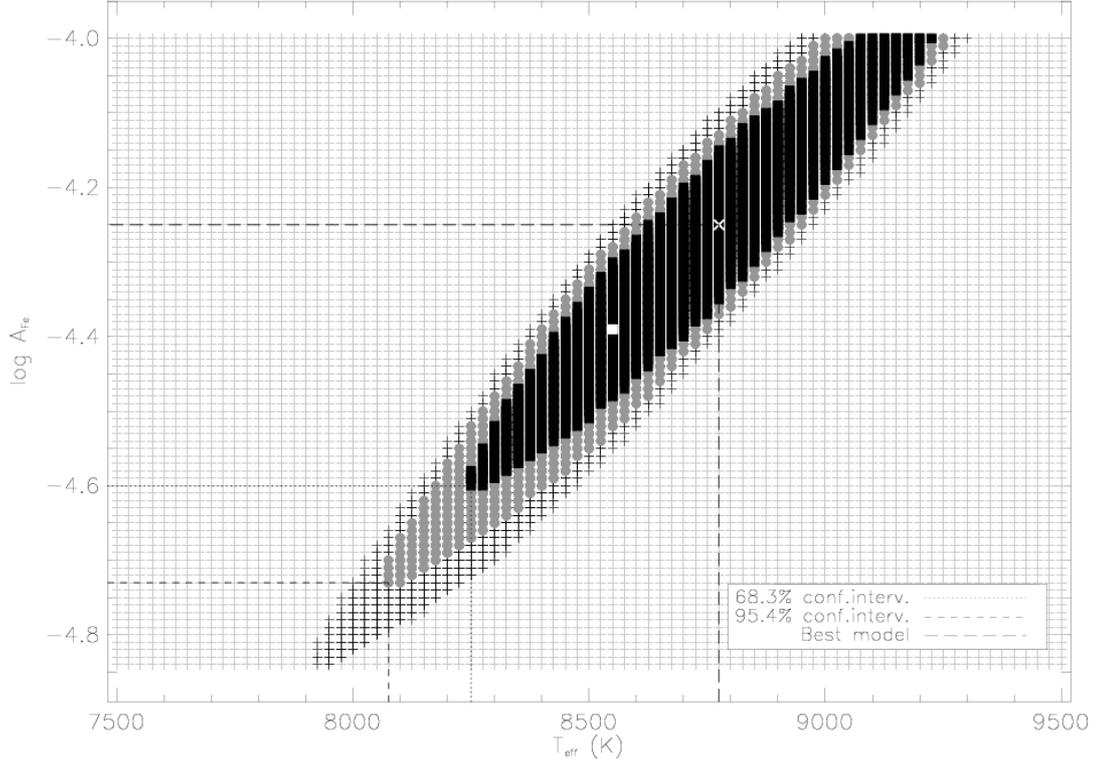}
\caption{2D-repartition of the $S_{\rm{red}}$ values as a function of \teff\ and $\log{A}_{\rm{Fe}}$. This graph has been built by considering successively every value of \logg, from 3.5 to 4.5 by step of 0.1. The white X represents the best model, whose \teff$=8775$\,K, $\log{A}_{\rm{Fe}}=-4.25$ and \logg$=4.2$. The boundary 68.3\%, 95.4\% and 99.7\% confidence intervals are respectively marked by the dotted lines and the black filled squares, the dashed lines and the grey filled circles and the black crosses. The grey cross represent the \teff\ and $\log{A}_{\rm{Fe}}$ ranges investigated in the present study from error box found in the literature, keeping in mind that \logg\ values range from 3.5 to 4.5. The white filled square locates the best model with the additional excitation equilibrium constraint described in the continuation of the Sect.~\ref{determin}.}
\label{figellipse}
\end{figure*}

A preliminary computation of $S_{\rm{red}}$ whithin the whole 3D-grid of synthetic spectra, whithout any constraint, yielded a \teff\ value of 8775\ K, a \logg\ value of 4.2 and a log A$_{\rm{Fe}}$ value of 
$-4.25$\,, i.e. [Fe/H]$=+0.29$\,   (the solar abundance being $\log{\left(\frac{N_{{\rm{Fe}}}}{N_{{\rm{H}}}}\right)_{\odot}}=-4.54$\,). The corresponding error bars were estimated using the method described above. For a confidence level of 68.3\%, we found the following uncertainties: 525\,K on \teff, 0.7 on \logg\ and 0.35 on log A$_{\rm{Fe}}$.  For a confidence level of 95.4\%, we determined the uncertainties of 700\,K on \teff\ and 0.48 on log A$_{\rm{Fe}}$, the small sensitivity to \logg\ does not allow the determination of the 95.4\% confidence level for \logg\ within our parameter space. The results are shown in Fig.~\ref{figellipse}.

\begin{figure*}[!t]
\centering
\includegraphics[width=150mm]{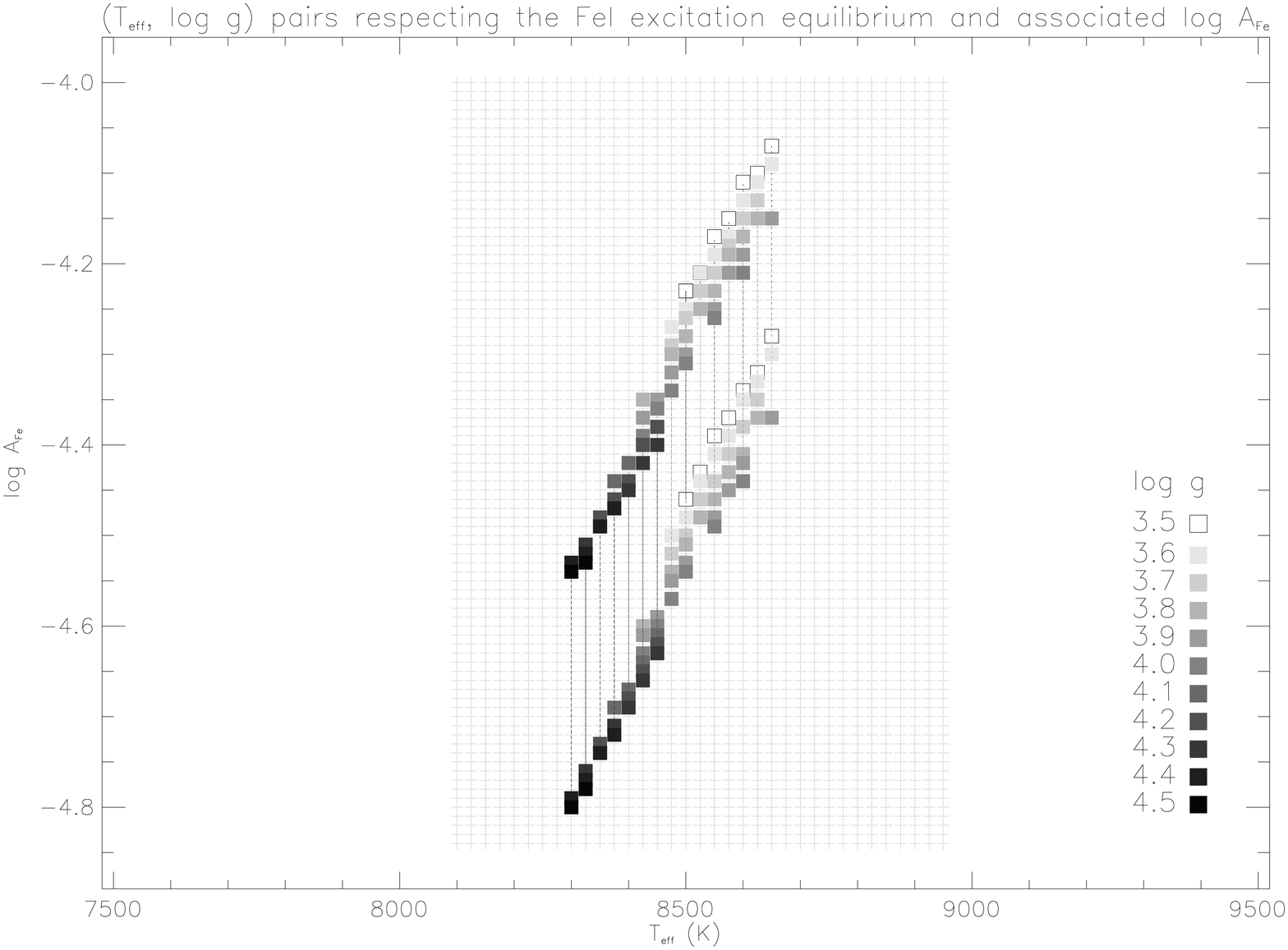}
\caption{(\teff, \logg) pairs respecting the Fe\,I excitation equilibrium and associated log A$_{\rm{Fe}}$ ranges (dotted lines delimited by squares). Each color of square, from lightest grey to darkest grey, represents a value of \logg\ from 3.5 to 4.5. Grey crosses represent the subspace of models for which there exists an intersection between the EW$_{\rm{obs}}$ line and the curve of growth for every one of the 8 selected Fe\,I lines.}
\label{figexceq}
\end{figure*}

What we pointed out at this stage were the limits of this first statistical computation due to the very low number of drastically selected lines used in the merit function minimization, that led to a significant \teff\ and \logg\ degeneracy: numerous synthetic spectra yielded $S_{\rm{red}}$ values within the chosen confidence intervals, and a much larger number of selected lines would have been necessary to converge reliably towards the model best representing the observations with smaller error bars. This effect was also pointed out by the results of a jackknife resampling, which consisted in recomputing several times the merit function $S_{\rm{red}}$, leaving out one after each other the equivalent width of one line.
This statistical method revealed that the \emph{best parameters} could change depending on the removed line within a $\Delta$\teff\ of $^{+150\,\rm{K}}_{-200\,\rm{K}}$, a $\Delta$\logg\ of $^{+0.2}_{-0.3}$ and a $\Delta$log A$_{\rm{Fe}}$ of $^{+0.10}_{-0.09}$. This line-dependance of the result was attributed to the lack of selected lines too.
To compensate for this lack, we confronted our preliminary result to the additional physical constraint brought by the excitation equilibrium: 
indeed, if  \teff\ is well determined, all spectral lines corresponding to a specified ion, and with a given excitation potential, should provide the same abundance determination. In a first step, for consistency, we only kept stellar model couples (\teff, \logg) for which the equivalent width EW of all selected Fe\,I lines had an intersection with the curve of growth (EW$_{\rm{synth}}$ as a function of $\log{A}_{\rm{Fe}}$) within the range of our parameter space. In a second step we calculated for each selected model linear regression lines through the 8 data points  $\log{A}_{\rm{Fe}}=f(\chi)$ ($\chi$ being the excitation potential, usually expressed in eV) and kept only the stellar models 
yielding a close to zero slope regression line (in fact we calculated a standard deviation of the distribution of all slopes, and kept models with slopes $0.\,\pm 1\,\sigma$). Fig.~\ref{figexceq} shows the fact that for each \teff ~a small range of  \logg\ and log A$_{\rm{Fe}}$ couples agreed with the excitation equilibrium within the error bar. 

\begin{figure*}[!h]
\centering
\includegraphics[width=150mm]{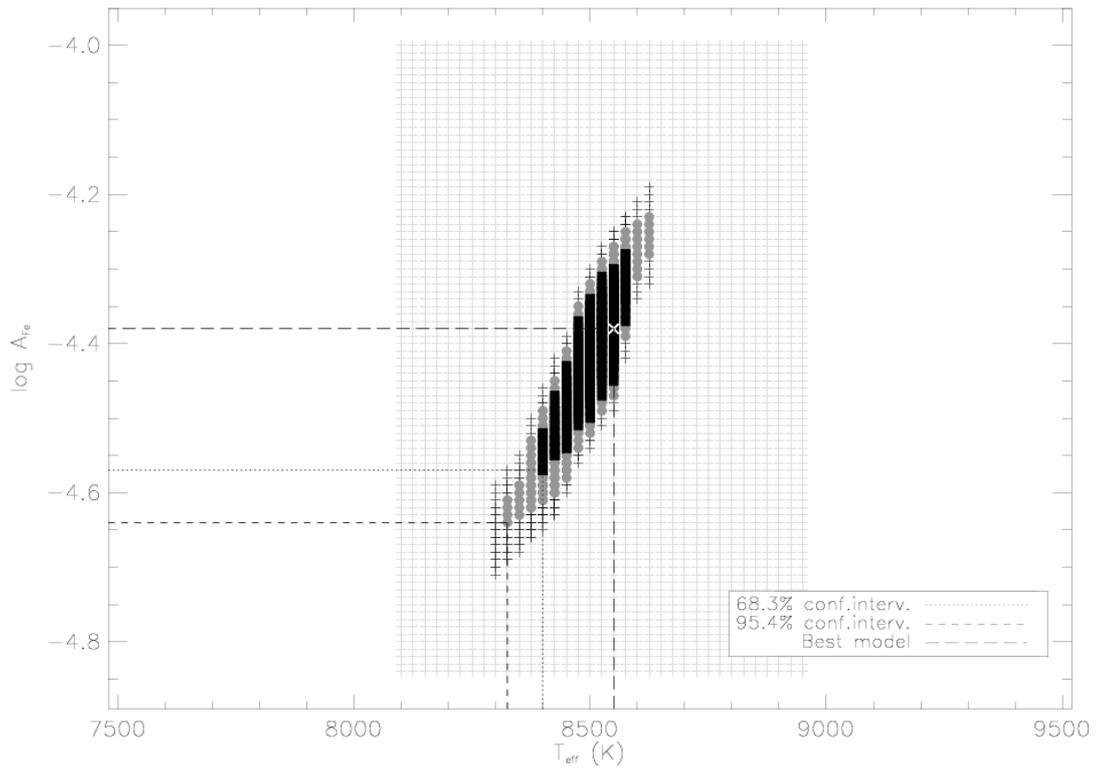}
\caption{2D-repartition of the $S_{\rm{red}}$ values as a function of \teff\ and $\log{A}_{\rm{Fe}}$. The white X represents the best model, whose \teff$=8550$\,K, $\log{A}_{\rm{Fe}}=-4.38$ and \logg$=3.9$. Confidence interval symbols are the same as those used in Fig.\ref{figellipse}. Grey crosses have the same meaning as in Fig.~\ref{figexceq}.}
\label{figelexceq}
\end{figure*}

Taking the excitation equilibrium into consideration, we excluded models not respecting it and searched again for the minimum value of $S_{\rm{red}}$. We obtained: \teff$=8550$\,K, \logg$=3.9$ and log A$_{\rm{Fe}}=-4.38\,$. To determine the error bar on these parameters, we adopted the same approach as described previously in this section. It should be noted that \logg\ is the less accurately determined stellar parameter because of the fainter sensibility of the selected lines to this parameter. The results can be seen in Fig.~\ref{figelexceq} and are summarized in Table~\ref{tablesum}. The shape of the 2D-repartition od the $S_{\rm{red}}$ values showed in Fig.~\ref{figelexceq} can be understood from Fig.~\ref{figellipse} and Fig.~\ref{figexceq}.

\begin{table}[!h]
\centering
\caption{Fundamental parameters of HD\,104237 determined in this study.}\label{tablesum}
\begin{tabular}{l|l|l|l}
\hline
 stellar & best model & 68.3\% & 95.4\% \\
 param. & param. values & conf. int. & conf. int. \\
\hline \hline
$T_{\rm{eff}}$ (K ) & 8550 & $\pm150$ & $\pm225$  \\
log $g$ & 3.9 & $\pm0.3$ & $\pm0.4$  \\ 
 & & & \\
log A$_{\rm{Fe}}$ () & -\,4.38& $\pm0.19$ & $\pm0.26$ \\
$\rm{[Fe/H]}$ () & +0.16 & $\pm0.19$& $\pm0.26$\\
\hline
\end{tabular}
\end{table}

\begin{figure}[!h]
\includegraphics[width=90mm,height=70mm]{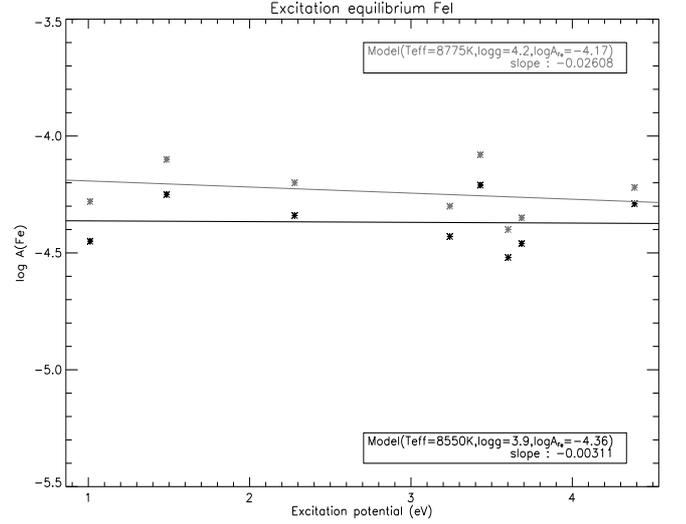}
\caption{Excitation equilibrium of Fe I. In black: the model minimizing $S_{\rm{red}}$ with excitation equilibrium constraint. In grey: the model minimizing $S_{\rm{red}}$ without any additional constraint.}
\label{figexceq2mod}
\end{figure}

In order to check the validity of our result, we plotted the abundances determined from the 8 Fe\,I selected lines as a function of their excitation potential in the case of the model minimizing $S_{\rm{red}}$ without any additional constraint and of the model minimizing $S_{\rm{red}}$ with excitation equilibrium constraint. The result can be seen in Fig.~\ref{figexceq2mod}. The linear regression across the points of the model with \teff$=8775$\,K shows a negative slope, whereas the one across the points of the model with \teff$=8550$\,K shoes a flat regression line, which confirmed this model as the best solution.

Another successful verification of the reliability of our results consisted in verifying the iron ionization equilibrium, i.e. that an optimal model must give the same abundances derived from Fe\,I and Fe\,II.

\begin{figure}[!h]
\includegraphics[width=90mm,height=60mm]{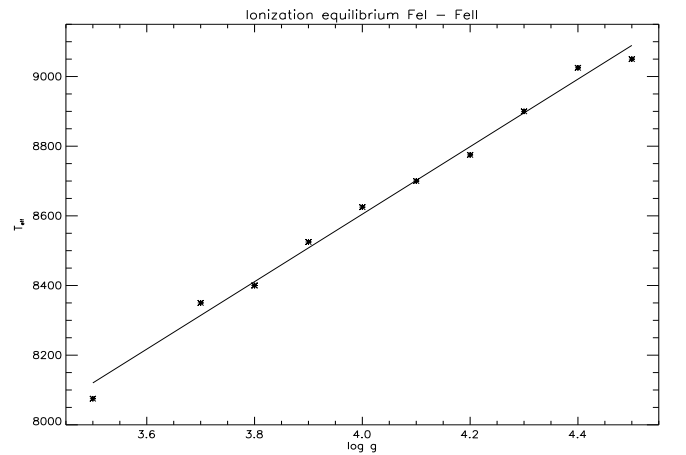}
\caption{\teff\ and \logg\ pairs respecting the ionization equilibrium between Fe\,I and Fe\,II lines in HD\,104237. Stars: \teff\ and \logg\ values derived from the HD\,104237 spectrum; the line shows a  linear regression.}
\label{figioneq}
\end{figure}

\section{Discussion and Conclusions}
\label{conc}

In this article we presented an in-depth analysis of the prototype Herbig Ae star HD\,104237. 
The star has been previously identified as a PMS $\delta$ Scuti pulsator, and the comprehensive line profile analysis
of the extended quasi-continuous high resolution spectroscopic time series obtained at SAAO from April 22$^{nd}$ - 26$^{th}$
1999 has confirmed the main frequencies discovered by radial velocity analysis in \cite{bohm2004}. For the first time, the less than
1.5\,$\%$ continuum level variations of the equivalent photospheric LSD profiles were studied and a direct confirmation of the presence of
at least one non radial pulsations could be performed.  Based on a Fourier 2D analysis we identified the dominant
mode as an azimuthal order m = $\pm$ 1, and a most likely $\ell$ value of 1 or 2. Since HD\,104237 is a moderate rotator with
\vsini~ of only 12$\pm$2 \kms it will be difficult getting access to additional mode identifications in the future, a potential approach
would be to organize multi-site continuous observations with significantly higher resolution echelle spectrographs, ideally with R $\geq$ 70000.
On the mode identification aspect a modification of the FAMIAS code working with asymmetric photospheric profiles would be a great progress. 

In order to prepare a forthcoming asteroseismic modeling of this particular star we needed to determine very precisely its fundamental stellar parameters. Since HD\,104237 is a multiple system with a nearby spectroscopic binary companion, we decided to concentrate on the 
periastron night of 12$^{th}$ of April 2000, present in our data set of spectra acquired at SAAO (and described in  \citealt{bohm2004}).
To combine all spectra of this particular night to one high SNR reference spectrum, we developed an optimized spectral continuum determination tool. As a result, the reference spectrum had a SNR value of close to 400 per resolved element at 550\,nm. 

The detailed study of the fundamental stellar parameters has provided values of $\teff = 8550 \pm 150$\,K, $\logg = 3.9 \pm 0.3$ and log A$_{\rm{Fe}} = 4.38 \pm 0.19$  (i.e. [Fe/H]= +0.16 $\pm$ 0.19), the error bars corresponding to the 68.3$\%$ confidence interval. A particular effort was put on the statistically correct determination of the associated error bars. In a review of \teff\ and \logg\ determination methods, \cite{smalley2005} mentioned optimal error bars to be of $\pm100$\,K for \teff, $\pm$0.2\, for \logg\ and of the order of 0.05 to 0.1\, in abundance. Due the high level of activity of HAeBe stars, and in the case our target star the spectroscopic binarity,  it is very difficult to obtain a sufficiently large number of unpolluted, exploitable photospheric spectral lines of a given ion; the achievable uncertainties remain therefore rather large.
In addition, a wrong estimate of the contribution of the secondary, both in spectral class and in luminosity, might lead to systematic errors which are very difficult to take into account. However, we estimate these errors to be of second order in our study. 

\begin{figure}
\includegraphics[width=90mm]{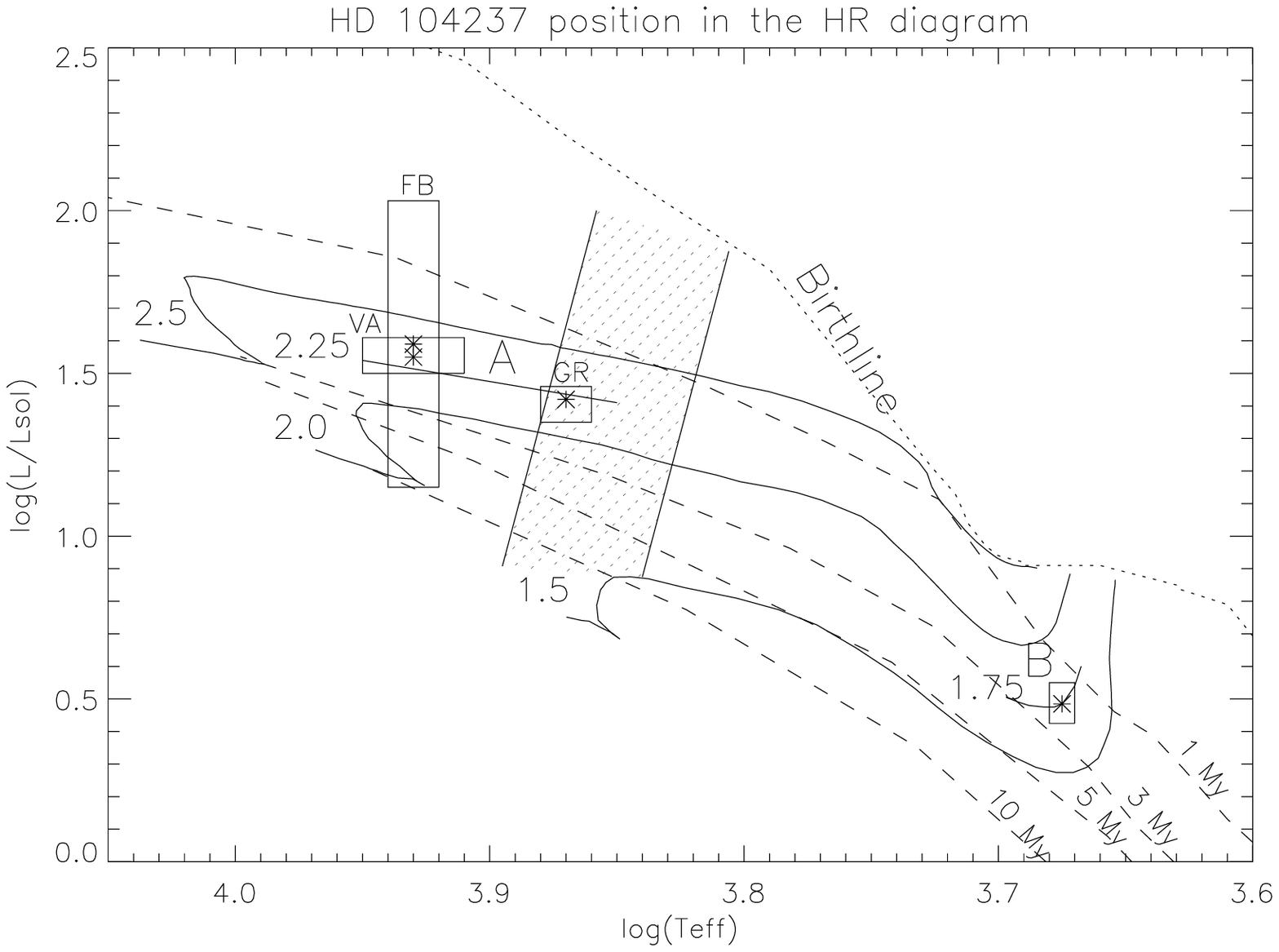}
\caption{New fundamental stellar parameters of the primary (A) component of HD\,104237, located in the HR diagram from \cite{bohm2004}.  The location of the secondary component HD\,104237\,b (B) is also indicated. The luminosity and effective temperature pairs reported in  \cite{vdancker1998} (VA) and  \cite{grady2004}(GR) are shown in the diagram,
including their {\it estimated} error bars, along with our new parameters (FB), associated with error bars corresponding to the 68.3$\%$ confidence interval. Evolutionary tracks (1.5,2.0 and 2.5 M$_{\odot}$) and isochrones are by Palla \& Stahler (2001).  The Marconi \& Palla (1998) instability strip is represented by the shaded area}
\label{fighrd}
\end{figure}

The surface gravity being difficult to constrain in our spectroscopic approach, the associated luminosity ratio (based on the empirical mass-luminosity relation by \citealt{malkov2007}) was determined to ${\rm{log}}\ (L_{\star}/L_{\odot})$ = 1.59 $\pm$ 0.44, a value centered on the result by 
\cite{vdancker1998}, whose authors determined ${\rm{log}}\ (L_{\star}/L_{\odot}) = 1.55^{+0.06}_{-0.05}$. On the contrary, our spectroscopic effective temperature determination is significantly improved with $\teff = 8550 \pm 150$\,K, again centered on the central value of \cite{vdancker1998}, whose result was $\teff = 8500\pm 500$\,K. Our independent spectroscopic approach therefore tends to confirm the previously photometrically determined values by  \cite{vdancker1998}. Combining the two results would yield the following most likely fundamental parameters:
 ${\rm{log}}\ (L_{\star}/L_{\odot}) = 1.55^{+0.06}_{-0.05}$, corresponding to $\logg = 3.93 \pm 0.09$, and $\teff = 8550 \pm 150$\,K.
Fig.~\ref{fighrd} shows well the intersection between our new results and previously determined parameters from \cite{vdancker1998}.
At this stage it is important to mention that error bars by \cite{vdancker1998}  and  \cite{grady2004} might be largely underestimated. 

\cite{ackewaelkens2004} determined an iron abundance of  Fe ([Fe/H]=$+0.09\pm0.19$. Our new result of the abundance, [Fe/H]= +0.16 $\pm$ 0.19, confirms the conclusion that HD\,104237 might be slightly overabundant. \cite{vick2011} calculated recently self-consistent stellar evolution models in this mass range including atomic diffusion; they concluded that for mass loss ranges above 10$^{-12}\,$M$_{\odot} $yr$^{-1}$
no surface abundance anomalies are expected. HD\,104237 shows a very strong H$_{\alpha}$ line with a complex profile, which tends to indicate significantly highter mass loss rates, similar to other Herbig Ae stars showing typical mass loss rates of the order of 10$^{-7} - 10^{-8}\,$M$_{\odot} $yr$^{-1}$ (see e.g. \citealt{bouret1998}). Taking these new simulations into account, we conclude on a close to solar iron abundance for HD\,104237, which is clearly within the error bars of both studies.

We want to mention at this stage that several assumptions have been made in this study, on the modeling side (LTE, plane-parrallel spectrum synthesis), on the contribution of the faint secondary (ratio of planck functions), on the threshold of line selection, but also the assumed photospheric line formation region of our lines of interest, amongst others. Despite the fact that these residual errors should be negligible with respect to our study, further improvements on line selection and analysis, such as a combined treatment of blended lines, will lead to more and more constrained fundamental parameters in the future.

A forthcoming article \citep{fumel2011} will describe the asteroseismic modeling of the primary component of HD\,104237.

\begin{acknowledgements}
The authors want to thanck N. Grevesse for interesting and fruitful discussions about stellar parameters determination and his
critical advices about atomic parameters.  Warm thanks also to H. Carfantan for precious discussions about the statistically meaningful use of the $\chi^2$ quantity and his advices concerning error bar determination in multiple dimension parameter spaces. Many thanks  also to 
F. Ligni\`eres, P. Petit and P. Fouqu\'{e} for their important comments on the approach we adopted. We acknowledge the Vienna Atomic Line Data base we used for our research work. Our thoughts are going to P. Reegen for his important contribution to time series analysis. 
\end{acknowledgements}

\bibliographystyle{aa}
\bibliography{fubotorsten24juin}

\end{document}